\documentclass[acmsmall]{acmart}
\usepackage{array}
\usepackage{multirow}
\usepackage{longtable}
\usepackage{soul}

\AtBeginDocument{%
  \providecommand\BibTeX{{%
    \normalfont B\kern-0.5em{\scshape i\kern-0.25em b}\kern-0.8em\TeX}}}

\setcopyright{acmcopyright}
\copyrightyear{202x}
\acmYear{202x}
\acmDOI{<DOI>}

\begin{document}

\title{Risk and Threat Mitigation Techniques in Internet of Things (IoT) Environments: A Survey}

\author{Marwa Salayma}

\email{m.salayma@imperial.ac.uk}
\orcid{0000-0001-5359-1455}
\affiliation{%
  \institution{Imperial College London}
  \streetaddress{180 Queen's Gate, Huxley Building, South Kensington Campus}
  \city{London}
  \state{England}
  \country{UK}
  \postcode{SW7 2AZ}
}

\begin{abstract}
Security in the Internet of Things (IoT) remains a predominant area of concern. This survey updates the state of the art covered in previous surveys and focuses on defending against threats rather than on the threats alone. This area is less extensively covered by other surveys and warrants particular attention. A life-cycle approach is adopted, articulated to form a ``defence in depth'' strategy against malicious actors compromising an IoT network laterally within it and from it. This study highlights the challenges of each mitigation step, emphasises novel perspectives, and reconnects the discussed mitigation steps to the ground principles they seek to implement.

\end{abstract}



\maketitle 

\section{Introduction} 
\label{sec:Intro}
Although we refer to \textit{the} Internet of Things (IoT) as if it formed a system, this terminology is not entirely correct and does reflect modern developments. Precursors of the IoT enjoyed various names such as Pervasive Computing, Ubiquitous Computing or Machine-to-Machine Communications (the latter taking a more network centric view) and sought to convey the idea that computation could occur anywhere. Indeed, the IoT is not just a collection of ``things'', nor a well defined system formed of things, but the instrumentation of the entire physical space surrounding us with an Internet connected digital interface and computational capabilities that increasingly comprise decision making and even learning. We often talk of \textit{"smart"} things whether it is, for example,  smart-meters, smart-buildings or, smart-toys. Again this only reflects that all objects that we are accustomed to in our physical spaces now comprise a digital component able to perceive the physical world through sensors and to control it through actuators. This point is particularly important when it comes to security. Compromising the security of the digital interface of a physical object impacts its physical behaviour and security, and the threats to be considered do not all originate in the digital (cyber) space but may start by exploiting their physical vulnerabilities or the trusting nature of their human users. Having made this point, this paper adopts commonly accepted terminology and refers to IoT Systems, Devices, Networks or Environments, bearing in mind that it is only a digital (cyber) perspective on the entirety of the physical world that surrounds us, which interconnects the physical world to the resources of the digital space.    

The number of IoT devices is continuously increasing, and this trend is set to continue. In their latest report, IoT Analytics estimated that in 2022, the global number of connected IoT devices grew to 14.4 billion, which is a 18\% increase compared to 2021, and by 2025, IoT Analytics predicts that there likely to be around 27 billion IoT connections \cite{IoTAnalytics}. In some respects, this may turn out to be an underestimate. On one hand the size of the devices is continuously reducing as well as their power consumption. On the other the (wireless) network connectivity is increasing e.g., the deployment of 5G  \cite{wasicek2020future}. Finally, devices are increasingly capable of learning and autonomous decision making. These trends will lead to more devices being used to monitor the physical world at a finer level of granularity and provide increasingly complex systems that optimise our usage of resources, personalise the services that are offered to us and, hopefully, increase our quality of life.    
However, adding an IoT device to a system is also adding an opportunity to compromise that system for a malicious actor. Any device connected to the Internet can be attacked from any other Internet location. Furthermore, in contrast to traditional computers or cloud servers securely hosted in offices or secure physical locations, IoT devices are deployed in the physical environment and can also be subjected to direct connections and physical attacks. Considering their vulnerability, a direct consequence of adding many IoT devices to our systems is that the attack surface of the IoT systems is also increasing exponentially. Faster interconnections, and rapid response also mean that compromises can spread faster and wider within the systems making them more difficult to protect and dependent on rapid response to a compromise to maintain their resilience. As well as making systems more robust to adversarial threats "by design", it is also necessary to deploy response techniques that can hinder the progress of an attack as well as responses that enable an adaptation (re-configuration) of the system and its recovery to maintain the system's function even when the systems have been partially compromised.  

The security and resilience of IoT environments is a complex topic that spans across the entirety of their life-cycle from design and realisation to their deployment, operation and decommissioning. In contrast to the other related surveys which  fall short of outlining concrete coherent steps to mitigate the spread of attacks in an IoT environment, this survey explores security measures in IoT system that are applied throughout the life-cycle of the IoT devices starting from their design, to the moment when a device joins a network, while the IoT device operates in the network, and until it leaves (or is /removed)  from the network and decommissioned. The survey discusses threat mitigation techniques applied across different IoT application contexts, and elaborates on how to apply each mitigation technique, its benefits and limitations and the extent to which progress has been reported in the literature.  In essence, the discussed measures provide answers to the  questions of how IoT device(s) connect and communicate with new devices and systems safely, starting from the moment when the IoT device(s) join the new environment, whilst operating in it, when an attack occurs, and until the device is removed/decommissioned or leaves the environment. This paper adopts a " defence in depth" strategy in discussing mitigation techniques proposed for the aim of controlling and slowing down the spread of threats in the IoT environment throughout the life-cycle of the IoT device. A defence in depth strategy to securing systems uses measures that aim to reduce systems vulnerabilities, contain threats, and mitigate attack effects if they occur, such that if an attacker manages to overcome one layer of defence, they still need to overcome the subsequent defence layers to compromise the system \cite{vacca2012computer}. The challenges are being addressed in the design of individual devices and in the design and operation of deployments. Like in the case of enterprise or more traditional computing environments, new techniques are being developed to make devices more trustworthy and new techniques are being developed to make systems more resilient and trustworthy by detecting, mitigating and responding to threats at run-time. 

After summarising the aspects covered in prior surveys in Section \ref{sec:prior}, we discuss aspects of self-protection and self-defense in Section \ref{sec:selfprotection}, techniques based on mediation in Section \ref{sec:mediation}, segmentation techniques in Section \ref{sec:segmentation} and techniques based on redundancy and recovery in Section \ref{sec:availability}, before drawing the conclusions in Section \ref{sec:conclusions}.

\section{Prior Surveys}
\label{sec:prior}
\begin{figure}[htp] 
    \centering
    \includegraphics[width=\textwidth]{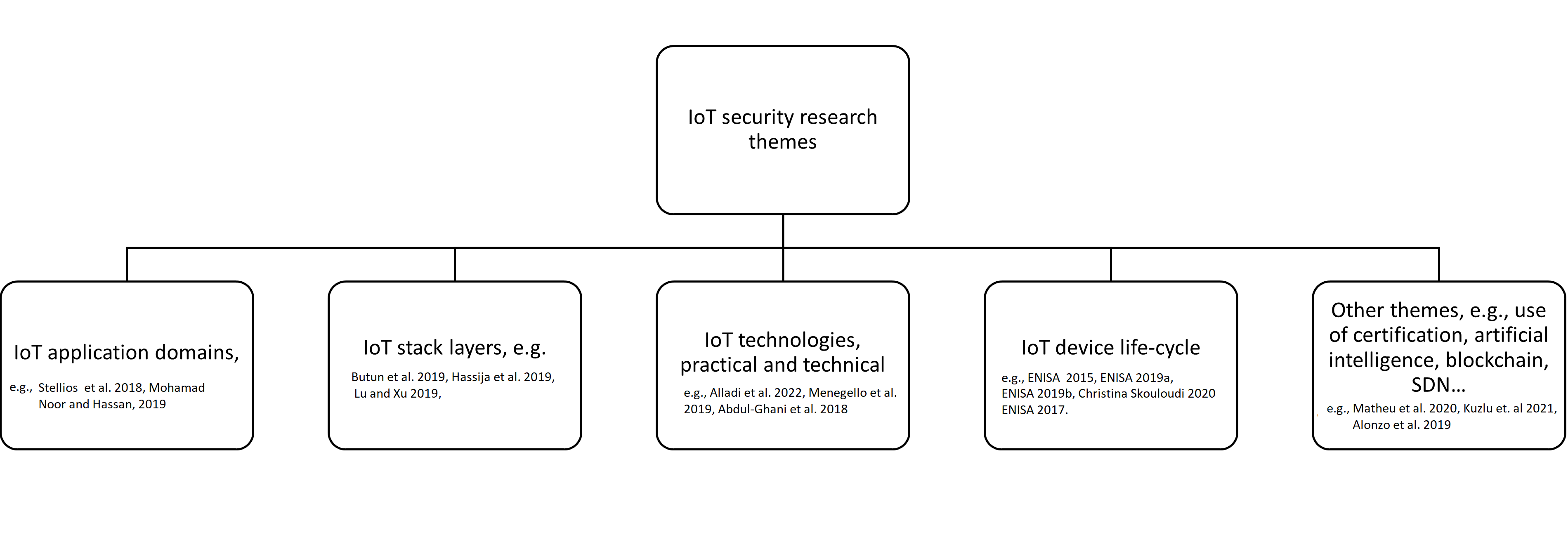}
    \caption{Research themes of the surveys on security in IoT environment}
    \label{fig:taxonomy}
\end{figure}

Many surveys have been published on IoT threats, attacks and countermeasures \cite{8742551, alladi2020consumer, hamza2020iot, matheu2020survey, meneghello2019iot, hassan2019current, butun2019security, stellios2018survey, abdul2018comprehensive}, at least twelve have been published very recently, between 2021-2023 \cite{stergiou2023security, kamalov2023internet, 10.1007/s11277-021-09420-0, kumar2023towards, hromada2023security, gerodimos2023iot, liu2023survey, abdel2022internet, rayes2022internet, Najmi2021ASO, CHOO2021102136,kuzlu2021role}. They introduce different classifications of IoT security challenges from varying perspectives. Examples of approaches taken by surveys are first described, before a figure representing themes is shown in Fig.\ref{fig:taxonomy}.

\citeauthor{stellios2018survey} \cite{stellios2018survey} classify IoT attacks as well as mitigation techniques across different application domains from 2010 until 2018. The attacks discussed comprise real-world incidents, as well as attacks that have been implemented and published as proof-of-concept, both are referred to in the survey as ``verified attacks''. The application domains considered include industrial control systems, smart power grids, intelligent transportation systems, and medical applications. The authors highlight the potential impact of attacks on critical systems when IoT devices are connected to them directly or indirectly. The IoT attacks considered include those that occur even when no IoT devices  connected to the critical infrastructure. Such hidden paths of attack are termed ``subliminal attack paths'' by the authors. The authors also proposes a risk assessment model consisting of threat, vulnerability and impact, based on which mitigation strategies are classified for each application domain. The study maps each of the security controls to the threats, and suggests who should be responsible for applying those controls, e.g., the owner, the administrator, the IoT manufacturer or the the regulator. The survey sheds light on the gaps in current security controls applied in each sector, and emphasises the inadequate implementation of current security controls owing to a lack of regulation and security policies that would force operators to use security tested, but usually more expensive IoT devices.

\citeauthor{MOHAMADNOOR2019283} \cite{MOHAMADNOOR2019283} discuss in a slightly earlier paper the research trends in IoT security and IoT security control strategies (mainly focussing on 2016-2018). The security controls discussed are applied according to the threat vectors. The authors observe that authentication is the most popular method at the application layer (60\%) followed by access control mechanisms, that involve trust evaluation. The authors observe that trends in the development of IoT security controls mainly focus on improving lightweight authentication and encryption for power and resource constrained devices. The authors also discuss secure routing, trust management, and the use of emerging technologies such as Software Defined Networks (SDN) and Blockchain. The survey concludes that IoT security mitigation should target all architecture layers, including perception, network, and application, whereas most of approaches focus primarily on the network layer. 

The surveys in \cite{alladi2020consumer, meneghello2019iot, abdul2018comprehensive} classify IoT security risks and mitigation techniques from a practical, technical, consumer and  application perspective and focus on the practical implications of IoT security. For example, \cite{meneghello2019iot} discusses IoT security from a more practical perspective compared to others and focuses on the security controls adopted in popular IoT communication protocols, such as: ZigBee, Bluetooth Low Energy (BLE), 6LoWPAN, and LoRaWAN, whilst highlighting the weaknesses of these controls. The survey also discusses other security mechanisms including the use of encryption, both standard and light-weight, random number generation, secure hardware, and Intrusion Detection Systems (IDS).  It presents a taxonomy of controls based on the different operational levels: information, access and functional, as well as a taxonomy of possible attacks at the different layers of the communication stack. The survey points out the importance of security by design and the systematic use of standard security mechanisms, which are often poorly implemented due to the heterogeneity of IoT devices. 

Other surveys \cite{butun2019security, 8742551, 8462745} also adopt a taxonomy for IoT threats and countermeasures distinguishing between the sensing, network and application layers. For example, \cite{8462745}, proposes a four layers IoT architecture: sensing, network, middleware, and application, and discusses potential attacks and threats that target each layer. The survey classifies IoT attacks into eight categories, and briefly introduces common countermeasures that apply not only to specific layers but also to ``intelligent'' objects and the entire network. The work discusses RFID-Based Authentication Measures, as well as measures that apply in Wireless Sensor Networks (WSN) and identifies a number of trends and emergent developments towards more secure IoT systems including: cloud service security, 5G, Quality of Service-Based Design, IoT forensics, self-management, and Blockchain Embedded Cybersecurity Design.

Although the taxonomies proposed in the surveys mentioned above provide a good summary of the work done so far in this space, and adopt different perspectives on IoT cybersecurity, they fall short of outlining concrete coherent steps to mitigate the spread of attacks in an IoT environment. In fact, threat mitigation strategies in IoT system should be considered and applied throughout the life-cycle of the IoT devices starting from their design, to the moment when a device joins a network, while the IoT device operates in the network, and until it leaves/removed from the network and is eventually decommissioned. To address these aspects, the European Union Agency for Network and Information Security (ENISA) published good practices that can applied in an IoT environment and guidelines that apply to every step of a product's life-cycle: its development, its integration in the system, and its usage and maintenance until end-of-life. The first guidelines were published in 2015 and targeted the IoT product life-cycle in the context of smart home environments \cite{ENISA2015}. More recently, in 2020, ENISA published guidelines on securing the IoT supply chain, that discuss the entire lifespan of IoT devices: from requirements and design, to end use, delivery maintenance, and disposal, recommending security measures for each step \cite{ENISA2020}. This work also builds on guidelines proposed by ENISA in 2019 \cite{ENISA2019}. With these studies, ENISA complements their baseline security recommendations for IoT \cite{ENISA2017}, which are combined in one tool for securing different ``smart environments'', such as smart hospitals, smart airports, smart cars and smart cities \cite{ENISA2019b}. Much of the state of the art established by ENISA is based on surveys and interviews with cybersecurity experts, IoT devices manufacturers, network operators, and standards groups.

Although the guidelines proposed by ENISA are comprehensive, they often assume that the user has control over the integration and usage of the IoT devices. In contrast, IoT networks often have to integrate devices that are outside of operator control. Another shortcoming is an insufficient consideration of the dynamic aspects of a system. In many cases, systems are composed of several devices that join and leave a network dynamically or that may be intermittently connected. Such devices are often mobile, either because they are mobile themselves e.g., autonomous vehicles, drones, or because they are instrumenting objects that are physically mobile e.g., body sensor networks for healthcare. This dynamicity requires security controls, such as authentication, access control, risk and trust evaluation and countermeasures to be applied continuously rather than at specific points in time. For example, the risk to a network can vary depending the type and number of devices joining the network. Finally, IoT environments can leverage redundancy to improve integrity and recovery. The same physical reality is often sensed through multiple sensors making it more difficult to compromise the perception of reality by compromising a single (or a few) devices. Redundancy in inter-connectivity permits to reconfigure networks in response to security events and isolate compromised devices, whilst enabling the rest of the network to operate normally. Finally, redundancy of functionality can help enable the recovery through adaptation and reconfiguration. But it is often not sufficient to consider the attacks alone. The impact of the attacks must be considered to determine the levels of redundancy required. This is particularly important in safety-critical systems and critical infrastructures.  

This study classifies prior surveys according to their focus, as shown in Fig.\ref{fig:taxonomy}. The survey distinguishes between a focus on: IoT application domains, stack layers, technologies, device life-cycle and other themes. Each one of these areas discusses main topics commonly covered in literature that relate to that theme. For example, studies vary in the number of layers considered, but most commonly discuss the: \textit{sensing}, \textit{networking}, and \textit{application layers} \cite{8462745}. 

Based on the presented related work, and their shortcomings, two questions are raised:
\begin{enumerate}
    \item How could IoT device(s) be enabled to, join, operate, and leave new environments in a secure way, such that the current IoT device(s) and systems in the environment stay protected from possible threats and attacks that might be caused by any newly joined infected device(s)?
    \item How can newly joined device(s) be protected from being infected or compromised by compromised devices that already exist in the network? 
\end{enumerate}

\begin{figure}[htp]
    \centering
    \includegraphics[width=10cm]{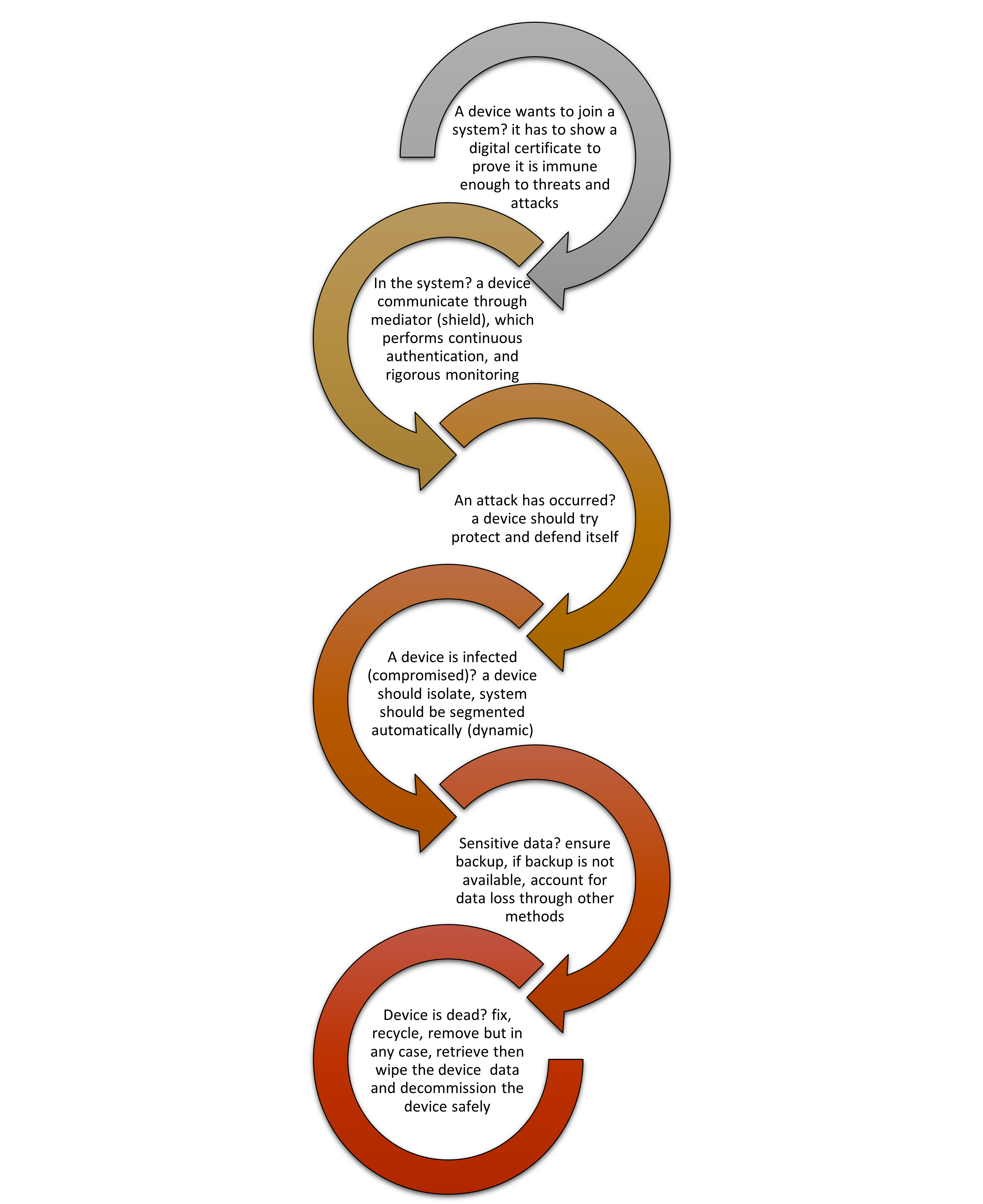}
    \caption{Mitigation techniques throughout the life-cycle of an IoT device: a defence in depth approach}
    \label{fig: fig2}
\end{figure}

To answer these questions, this study explores below additional security measures that can be applied across different IoT application contexts, and elaborates on how to apply each mitigation technique, its benefits and limitations and the extent to which progress has been reported in the literature. By contrast to the vast majority of the research directions presented in literature which is related to this work, this survey does not discuss the mitigation techniques in the IoT domain from the layered network architecture of an IoT system, such as IoT-Cloud/Edge, IoT-IoT, and IoT-gateway. In another words, the discussions in this work is not structured  based on each network layer of the IoT device or its associated protocols. This is because discussing the topic from the layering architecture perspective does not help to answer the two questions presented above: the layering discussion does not account for the dynamicity of the IoT environment, one of which is the mobility of the IoT device itself when joining or leaving a network, which is one of the main reasons for the fast propagation of a cyber attack in the IoT environment. Rather, this survey provides answers to the  questions of how IoT device(s) connect and communicate with new devices and systems safely, from the perspective of discussing the life cycle of the device(s), starting from the moment when the IoT device(s) join the new environment, whilst operating in it, when an attack occurs, and until the device is removed/decommissioned or leaves the environment, and hence the discussed measures address the dynamicity of the IoT environment. Those steps are  depicted in Fig \ref{fig: fig2}, which shows the mitigation steps as a continuous process run in a ``defence in depth'' approach starting when the device tries to join a system until it dies or leaves. Those mitigation techniques are summarised in Table \ref{tab: Table1} and are discussed in detail in the following Sections.   

\section{Device self-protection and self-defence}
\label{sec:selfprotection}
Securing IoT environments starts with securing the devices themselves. Given that IoT devices are typically exposed to both physical and cyber threats, a ``zero trust" assumption is often a safe default i.e., a device should not be connected unless it has been authenticated, authorised and reasonable steps can be taken to trust the device's integrity and even its capability to defend itself.  
If a device is able to prove its integrity and show its level of robustness to external threats, then it may be allowed to join a network without increasing substantially the risk to the network itself. Although some studies argue that such proofs could take the form of digital certificates, this only raises other questions: who it issue such certificates?, based on which verification processes?, what are the cost implications? Even then TOCTOU (time-of-check, time-of-use) issues remain i.e., how to ascertain that nothing has interfered with the device's integrity or trustworthiness since the last check. 
After securely joining the network by providing a certificate, software and hardware self protection and  defence mechanisms including adopting advanced techniques to make it even harder for the attacker to perform the attack,  all help the IoT device continue operating in the environment even after being infected from newly joined devices and vice versa. Those mechanisms are discussed in the following subsections. 

\subsection{Hardware self-protection and self-defence} 
\label{sec:HWhardening}

More emphasis is often put in the related work, on software and communication trustworthiness than on device hardware protection, although a number of research groups have focused on hardware aspects. The hardware of an IoT device can serve as a "root-of-trust" for subsequent integrity verifications, so ensuring its security is particularly important. It needs to be robust to physical attacks, supply-chain attacks and side-channel attacks amongst others. Physical attacks, include modification of the hardware circuit e.g., by removing the component's physical packaging and/or modifying the hardware. Supply-chain attacks refer to modification of the components in a hardware circuit or the circuit itself by suppliers (with or without their knowledge). Side-channel attacks refer to attacks conducted by monitoring or using the properties of the hardware such as power consumption, execution time, behaviour under faults \cite{1260985}.

An example of a physical attack is a Hardware Trojan (HT) -- a circuit inserted into a larger one to alter its function  and implement malicious behaviour. Hardware Trojans are discussed thoroughly in \cite{jsan8030042}, within the broader context of challenges of securing IoT hardware. \citeauthor{jsan8030042}  \cite{jsan8030042} consider HTs as the biggest hardware security threat, provide a taxonomy for HTs, and discuss specific mitigation and countermeasures for HT including: detection, Design for Trust (DFT) and split manufacturing. 

A number of methods are discussed in \cite{jsan8030042} to detect the modifications, which are classified into \textit{pre-silicon} i.e., validating the integrated circuit (IC) at the design stage, and \textit{post-silicon} i.e., verifying the fabricated IC at the manufacturing stage techniques. Post-silcon detection is further divided into \textit{destructive techniques} such as ``depackaging" the IC using reverse engineering techniques, and \textit{non-destructive techniques} such as verifying the fabricated IC using testing methods including functional tests, side channel analysis and automatic test-pattern generation (ATPG). \citeauthor{jsan8030042} also discuss other HT countermeasures to complement HT detection, as they advocate embedding HT prevention methods during the design phase to implement the concept of \textit{Design for Trust (DfT)} \cite{jsan8030042}. DfT involves methods to prevent HT insertion, such as obfuscation, layout filler, camouflaging, and use of trusted modules to monitor the IC at run-time and trigger security controls to mitigate the effects of HTs. However, despite its benefits, run-time monitoring incurs significant additional overhead in a resource environment that is already heavily constrained. ``Split manufacturing" stands as a third countermeasure against HT, aiming to hide the design intent of the IC to prevent malicious insertion \cite{jsan8030042}. 

Other hardware security controls seek to identify and authenticate devices uniquely. These include the use of Key Injection, Physically Uncloneable Functions (PUF) and the use of Hardware Security Modules (HSM) \cite{hamadeh2017area, lu2020xtseh}. HSMs as well as Trusted Platform Modules (TPMs), and more generally Trusted Computing plaforms seek to establish a hardware root of trust on the platform that can then be used to ascertain the platform's integrity e.g., through attestation, authenticate the device or securely perform cryptographic functions. 
  
However, TPMs (including TPM 2.0) are costly to use in a resource constrained devices, due to the additional space and power consumption they impose. To address this resource issues, the Trusted Computing Group (TCG) created Device Identifier Composition Engine (DICE), a security standard with lightweight hardware requirements, that can be used for hardware-based cryptographic device identity, data encryption, attestation of device firmware, and safe deployment. It can be implemented in a small size micro-controller, and hence is suitable to use in resource-constraint IoT devices. DICE has been first to be adopted by Microsoft for Azure IoTs. 

Another hardware based technique to provide security in the IoT context is the Remote Attestation (RA). RA presents a security technique in which a trusted verifier  assures the integrity of a prover i.e the  untrusted device. RA starts by the prover sending evidence about its current memory status to the verifier, which in its turn tries to match the received evidence with the expected state of the prover, according to which the verifier validates the trustworthiness of the prover \cite{seshadri2004swatt}. There are a number of research efforts that propose secure and lightweight hardware architectures such as  SMART \cite {eldefrawy2012smart} and TrustLite \cite{koeberl2014trustlite} to provide a secure remote attestation for embedded and IoT devices. Such architectures adopt minimal hardware components such as simple memory protection units (MPU) \cite{mohan2018special}. However, such architectures fail to provide secure attestation to large number of IoT devices such as  drones. 

Physical attacks cannot be mitigated without designing circuits that are tamper resistant. Different approaches have been proposed in the literature and are discussed in \cite{1260985}. These are divided into (i) attack prevention techniques, such as components packaging, and designing hardware circuits with independent power and timing properties, (ii) attack detection, such detecting illegal memory accesses by un-trusted software at run-time, (iii) attack recovery, for example displaying a security warning and rebooting the system, (iv) and tamper evident design. 

Tamper proof and tamper evident design can be used at different levels of complexity to meet different levels of physical security requirements from minimum protection (e.g., seals or enclosures) to environmental failure protection and testing, which is the highest level of protection. 

IBM’s 4758 PCI cryptographic adaptor is an example of adoption of such techniques, as the chip includes internal tamper circuits and sensor components, to detect and respond to physical penetrations, temperature and voltage attacks. However, applying such security levels of protection is expensive, far beyond the cost of typical IoT devices. 
To avoid physical attacks on the bus, \citeauthor{{1260985}}\cite{1260985} propose adopting processors that encrypt and decrypt all information sent on global buses and ensure that only encrypted data remains in the open memory, as well as encrypt data before it is sent outside I/O. This approach incurs high performance overheads, but also its implementations remain vulnerable to side-channel attacks, as also recognised in \cite{1260985}. 

Mitigation techniques against side-channel attacks are also discussed in \cite{1260985}. These include methods, such as randomisation, to reduce the system's exposure to monitoring and analysis of side-channel information, such as power, timing, and electromagnetic radiation. 

Randomisation is effective approach as it imposes a significant  extra burden on the attacker, and is not entirely unrelated to Moving Target Defence techniques employed at the higher layers.  

Other methods proposed by the authors against power analysis attacks, aim to increase the number of samples needed to conduct the attack by applying \textit{data masking}, introducing noise into the power measurement data and the use of reduced signal amplitudes. Suggested mechanisms against electromagnetic analysis attacks include applying aggressive shielding and techniques to break-out the layout of a chip by spreading the chip components across the chip surface. 

Methods for detecting \textit{fault injection attacks} and preventing transient fault attacks on cryptographic hardware are suggested to harness sensors that can monitor environmental properties to detect fault injection attacks, leveraging error detection methods to prevent such attacks. 
 
From another angle, hardware information has been proven to assist in malware detection e.g., by applying Machine Learning (ML) techniques to the low-level micro-architectural features captured by Hardware Performance Counters (HPC) with Machine Learning (ML). This has led to the so called \textit{Hardware-assisted Malware Detection (HMD)}, which has been shown to offer improvements on traditional software-based malware detection techniques \cite{ 9474036, 10.1145/2485922.2485970}. \citeauthor{10.1145/2485922.2485970} \cite{10.1145/2485922.2485970} were the first to harness HPC for malware detection. However, it was not long before evasion techniques were proposed. An adversarial attack on HMDs through which the malware detection accuracy is reduced is proposed in \cite{9474036}. In response, \citeauthor{9474036} \cite{9474036} aims to improve HMDs to be robust to adversarial attacks through Adversarial Training (AT).

\begin{figure}[htp]
    \centering
    \includegraphics[width=\textwidth, height=15cm ]{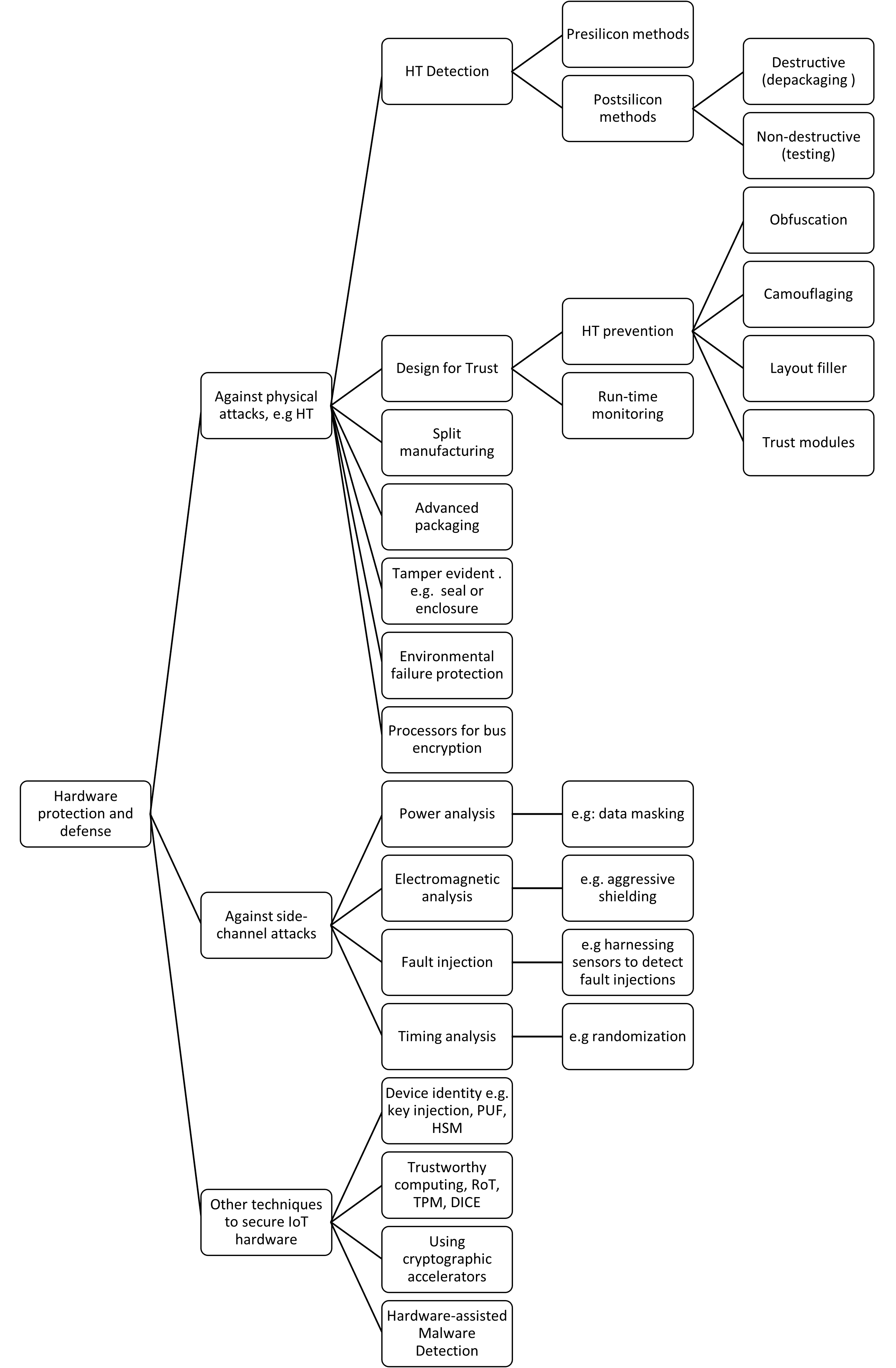}
    \caption{IoT device hardware protection and defence approaches}
    \label{fig: fig3}
\end{figure}

The IoT device hardware protection and defence approaches discussed in this section are summarised in Fig \ref{fig: fig3}.

Despite its benefits, hardware protection can have significant implications in terms of increased cost and increased overhead. For example, adding dedicated crypto-processors adds significant cost to the device, whilst monitoring techniques impose high system overhead. Protection against side-channel attacks also introduces significant overheads as the observability of the channel needs to be reduced e.g., by redundant or irregular use. The overhead is not only in terms of processing but also in terms of power consumption, reducing the applications in which the IoT devices can be used. Whilst effective, hardware protection can only be deployed in applications where the additional cost is justifiable and the additional overhead can be tolerated. This contrasts with the broad use and adoption of IoT devices that mainly leverages their reduced cost and power consumption. 

\subsection{Software protection}
\label{sec:SWhardening}
Software-based approaches to improve robustness against attacks are categorised in \cite{1260985} according to three design considerations: (i) ensuring the integrity and privacy of sensitive code and data at every stage of software execution, (ii) ensuring security when executing a given program, and (iii) removing software vulnerabilities that make the system vulnerable to attacks. Software attacks typically performed by malware exploit flaws in the program and its execution. Such flaws are vulnerabilities when they enable malicious actors to gain privileges in the system, access its content and control (at least partially) its execution \cite{1260985}. 

Software self-protection and self-defence approaches, can be two fold: (i) tamper-resistance techniques that deal with attacks when they occur, resist and counter their effect, and (ii) hardening techniques that aim to reduce the number of vulnerabilities. 

\citeauthor{1260985} \cite{1260985} discussed techniques to ensure software integrity such as adding hardware to support tamper resistance,  secure bootstrapping, enhancing OS security, ensuring integrity of software with safety checks, as well as software authentication and validation. They describe an example of secure bootstrapping tailored to the IBM PC architecture, which exploits the layered nature of the boot process, starting from turning on the system and moving from layer to layer verifying software integrity at each one.
 
Common approaches for hardware support rely on physical isolation, e.g. secure co-processors and memory subsystems for secure storage to which access is only allowed to trusted system components. Dedicated hardware for bus monitoring can also be used to protect the memory from illegal accesses. 
A bus monitoring system allows to detect illegal access to sensitive memory regions, and take suitable actions in return, such as zeroing memory areas. 

Methods to enhance the security of OS, such as applying strong process isolation and attestation are also discussed in \cite{1260985}. Process isolation involves protecting a process' private resources from one another, whilst attestation aims to guarantee the integrity of a process before it runs. For example, a common approach is to compute a hash code of the software and verify it against a pre-computed value before it runs. 
 
Process isolation can be achieved through sandboxing, restricted control transfers, and code origin checks. Other techniques aim to ensure that a process will not violate its security policies. These include proof-carrying code but also process shepherding, which monitors all control transfers with the aim at detecting and stopping malicious code from being executed. Programming language and software verification techniques can be used to verify the implementations and are commonly used for security protocols. They aim to detect software flaws (including vulnerabilities), ensure correctness and verify the implementation. Although such formal verification techniques have difficulty scaling to larger software implementations they can be more readily used in IoT devices where the code-base tends to be smaller. 

Software and systems hardening is defined by the National Institute of Standards and Technology (NIST) \cite{26021} as the process of eliminating reasons that could attract the attackers commit attacks by turning off non-essential services and patching vulnerabilities. It is particularly important in IoT environments especially when the devices are connected to the physical world (Cyber-Physical Systems) and are used in the context of decision making for the physical world. In such circumstances safety considerations also apply. Hardening techniques for IoT devices are discussed across several studies in the literature. For example, it is recommended as good practice for IoT manufacturers to increase the robustness of their products by adding safety algorithms to overcome accidental or intentional faults and decrease their impact. This is also applied in the context of cryptographic software. For example, \citeauthor{10.1007/978-3-030-64348-5_25}\cite{10.1007/978-3-030-64348-5_25} studied the effectiveness of five common software hardening techniques applied to a lightweight block cipher, called PRESENT aimed at resource constrained environments, such as WSN and RFID \cite{bogdanov2007present}. The hardening techniques considered include classic loop hardening, variable duplication, function duplication, decryption in place, and statement based counters, and were evaluated in preventing sensitive data leaks, realising their security and their impact on software performance. Their study revealed that redundancy hardening techniques on a functional level generally provide a good balance between fault tolerance, and software security, whereas classic techniques, such as  classic loop hardening, are more vulnerable. The study also indicates that, generally, hardening alone is not sufficient to mitigate faults, and needs to be combined with other techniques to increase software robustness. 

Systems hardening techniques are crucial to minimise attack surfaces that provide open doors or possibilities for attackers to launch attacks. Default factory usernames and passwords, and open ports are often scanned by malware to enable compromise; the Mirai botnet is only one example. 
An analysis of the Mirai botnet, and software hardening techniques to protect IoT devices against it are presented in \cite{frank2018protecting}. The study analyses Mirai malware behaviour when it searches and compromises new devices. The study conducted code auditing, discovering novel signatures in the communication between the Command and Control (C\&C) server and the bot for loading the Mirai malware, and compromising new bots, and rated and tested two hardening scripts. One proved successful in continuously detecting and  removing existing Mirai malware from an infected bot. The other proved successful in preventing the botnet from infecting a new device by making various configuration changes on the IoT device.
\citeauthor{Choi2018SystemHA}\cite{Choi2018SystemHA} also propose a framework that leverages system hardening and security monitoring to minimise vulnerabilities in IoT devices that do not implement security by design. The authors have analysed various IoT vulnerabilities defined by the Open Web Application Security Project (OWASP), and implemented a service prototype to detect malware in infected IoT devices. 
Hardening typically involves removing un-necessary functionality. However, in IoT devices, what is or is not essential is more difficult to predict at the outset, and \citeauthor{10.1007/978-3-030-64348-5_25}\cite{10.1007/978-3-030-64348-5_25} show that some software hardening techniques can negatively impact system software and its performance. Hardware components e.g., for tamper resistance can be effective but they add to the cost and the design of the device and also need to managed and used judiciously. 

\begin{figure}[htp]
    \centering
    \includegraphics[width=\textwidth, height=13cm ]{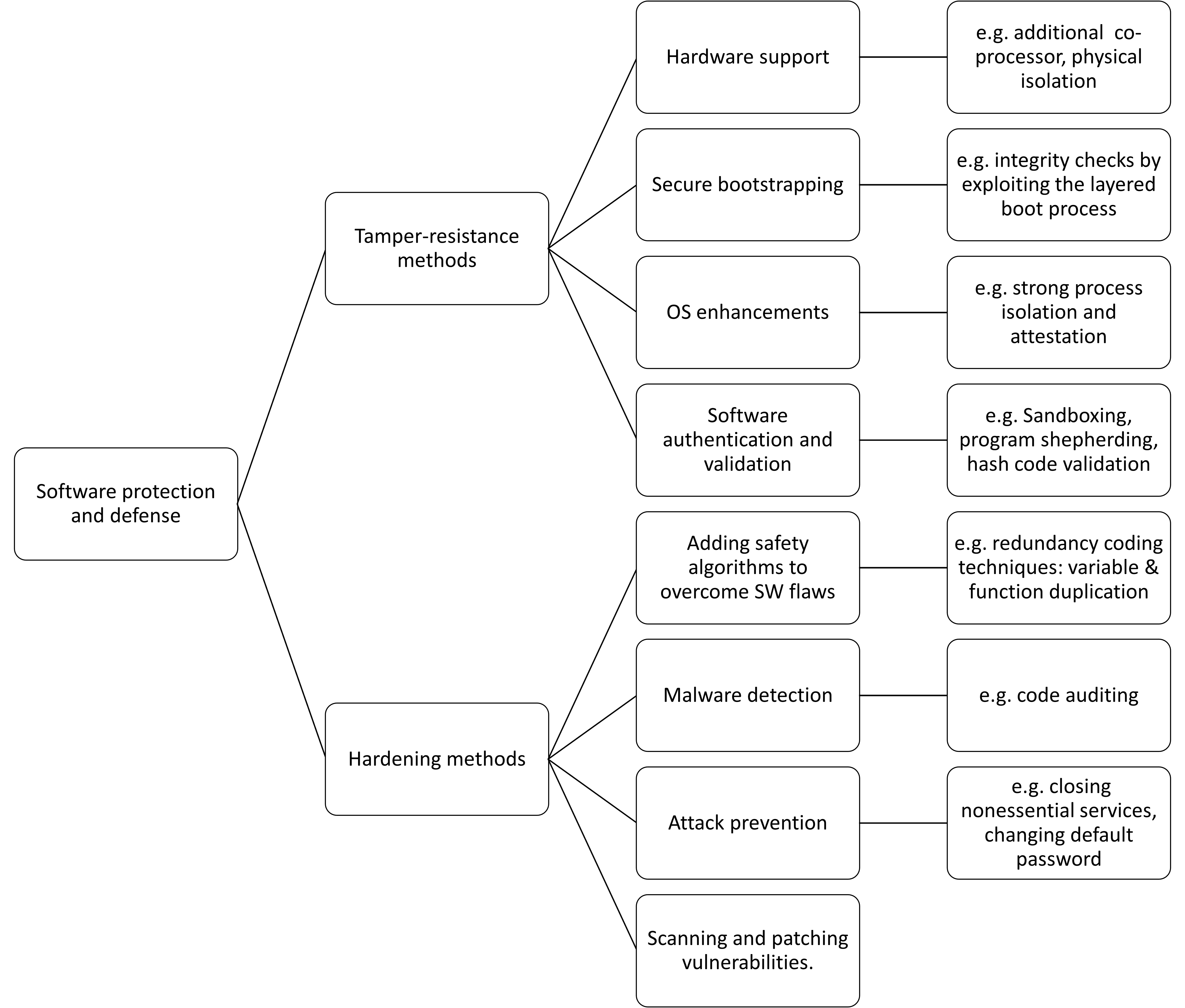}
    \caption{IoT device software protection and defence approaches}
    \label{fig:fig4}
\end{figure}

On another direction, the literature presents a number of software based RA techniques that do not rely on specialized hardware components. The survey in \cite{ankergaard2021state} discusses multiple software-based RA techniques,  the potential and challenges of applying them to provide security in the IoT context, for example it discussed methods to integrate RA with Blockchain to provide secure real time IoT and embedded systems (e.g., decentralization, traceability, anonymity and non-repudiation) such as Vehicle-to-Vehicle communications \cite{xu2018remote, fortino2020trust}.

Fig \ref{fig:fig4} summarises the main techniques mentioned so far in this section. These techniques can help considerably secure an IoT environment but often have high performance costs. Hardware based techniques can establish significant roots of trust, however this comes at a higher monetary cost to the realisation of the IoT device. Moreover, they also need to be managed.  

\subsection{Certification}
\label{sec:certification}

Certificates in the form of e.g., public or secret key certificates are a well adopted mechanism for providing a cryptographically signed proof that a certain process (e.g., verification of identity in the case of identity certificates has taken place). It would be relatively straightforward to use them to prove that a particular device has been through a certification process to prove that it has been checked for compliance with a number of requirements and has a certain level of robustness against attacks. However, the challenge is to define the process and the criteria for such certification.  
Cybersecurity digital certification is defined by the Committee on National Security Systems (CNSS) a “Comprehensive evaluation of an information system component that establishes the extent to which a particular design and implementation meets a set of specified security requirements.” \cite{CNSSI}. However, certification is not a trivial process, and certifying devices remains a significant challenge. In fact, there are many major questions that need to be answered in this arena, such as "who does the certification and what to certify", as discussed in \cite{8352084}. \citeauthor{8352084} \cite{8352084} highlight that both security and privacy might not be certifiable properties because ``IoT'' environments cannot be defined generically, so far there is even no clear definition for what IoT is. Moreover, there is no standardised methodology for certification, and due to the number of sectors and different environments in which IoT devices are used, defining generic certification processes would be a significant challenge. It may be possible to design such certification processes in specific sectors and for specific categories of devices, e.g., consumer IoT, however, the aim to make such certification processes generic conflicts with the need to have well specified requirements for certification as only features expected to be common can be certified. Certification is also usually a very costly and time consuming process (see for example certification of medical devices) thus delaying adoption and imposing upon the manufacturers singicant additional delays and costs in an ever evolving environment where the time-to-market is short and the profit margins are thin \cite{10.1145/3410160}.

\citeauthor{10.1145/3410160} \cite{10.1145/3410160} provide an overview of the main cybersecurity certification schemes and discuss the challenges of adopting them for the IoT. They further discuss the efforts related to the basic building blocks processes of cybersecurity certification, as proposed by the European Telecommunications Standards Institute (ETSI) \cite{ETSI}, which include the risk assessment and testing frameworks. The survey provides recommendations to enable applying cybersecurity certification to emerging technologies, such as the IoT, and propose a multidisciplinary IoT cybersecurity certification framework that integrates research and technical tools with policies and governance structures. Going further \citeauthor{8352084} \cite{8352084} suggest three approaches for certification: certifying products, certifying the processes that produce the products, and certifying the people who produced the product.
\citeauthor{matheu2019risk} \cite{matheu2019risk} propose an automated security certification methodology for IoT environments within a specific context. The framework is built on top of other security testing and risk assessment approaches tailored for IoT such as those proposed by ETSI. The work adopts the concept of a cybersecurity label, as a means to provide transparency on the outcome of the certification process. The label depicts information related to the level of security verified in the certification process. The methodology proposed was applied to a specific scenario that uses IoT protocols, such as the Constrained Application Protocol (CoAP). 

Certification for IoT can have significant benefits, but it remains significantly challenging. As stated in \cite{8352084} ``When it comes to dealing with the IoT certification quagmire, good enough is better than nothing, although good enough remains in the eye of the beholder.''

\subsection{Moving Target Defence (MTD)}
\label{subsec:mtd} 

System self-protection is crucial to improve the ``immunity'' of a device against threats, especially known threats, however, the static nature of the design of the software and the hardware makes it still vulnerable to attacks. For example, some IoT applications require numerous IoT devices to be deployed for environmental monitoring around the globe. Even if those devices are hardened before deployment, they will typically employ static and identical software configurations and hardware components. If an attacker succeeds in exploiting a vulnerability in one application of one device, the same vulnerability can be exploited in the many thousands of other devices that are deployed. Things get even worse as it takes in practice a long time to update and patch the software across a large number of devices, which means that the whole IoT system will be subject to attacks using that vulnerability for long time \cite{8333141}. 

\textit{Moving Target Defence (MTD)} has been proposed in recent years as a new type of defence that aims to leverage the attacker's lack of knowledge of the system topology and operation and aims to make it difficult for an attacker to perform the attack, by making the attack process exhausting and time consuming. To achieve this, MTD can be applied through three techniques, which all present a deterrent in space and time for the attacker to perform the attack. Those techniques are: (i) randomisation, or non-determinism in the internal system while ensuring it achieves same functionality, (ii) diversity, which refers to deploying non-homogeneous system components, to avoid a component from being breached by the same attack, (iii) dynamism, and refers to changing the system properties regularly so to hinder same attack from compromising it in the future. 

\citeauthor{8333141} \cite{8333141} describe these techniques in detail, and discuss possible methodologies to apply them with respect to the generic layered design of a computing system including the hardware, operating system, run-time environment, and data applications. Their chapter highlights the benefits, challenges and limitations of adopting MTD defense techniques in general. They propose three criteria to evaluate the effectiveness of applying MTD techniques: timeliness, unpredictability and coverage. Based on these criteria, \citeauthor{8333141} \cite{8333141} identify several issues that need to be addressed when implementing MTD in practice. For example, one needs to consider the balance between preserving the system's performance requirements and the cost incurred by applying MTD, especially in IoT systems. A very recent study \cite{9521488} targets specifically this trade-off, and proposes a framework that employs MTD by randomly shuffling the communication protocols between IoT nodes and an IoT gateway. The  strategy proposed aims to balance the increased overhead resulting from the use of an MTD approach and its impact on system availability with minimising the chance attack success. The proposed framework addresses the MTD design issues, which are: ‘‘What to move,’’ ‘‘When to move,’’ and ‘‘How to move’’ aiming to use the framework as a guideline for applying MTD in IoT environments. \citeauthor{navas:hal-02887462} \cite{navas:hal-02887462} go further proposing a generic modular MTD framework for IoT environments, that can be adapted to the requirements of a given IoT network. The framework adopts two MTD strategies; one that targets UDP port-hopping (having evaluated its effectiveness in a scenario where nodes were exposed to a remote Denial-of-Service attacks), and another that targets the Constrained Application Protocol (CoAP) resource URIs. 

\citeauthor{9270287} \cite{9270287} provide a systematic review of the existing IoT MTD techniques up to July 2020. The survey categorises the techniques based on entropy-related metrics and validates the suitability of the MTD techniques to improve the resilience of IoT systems. \citeauthor{9270287}\cite{9270287} conclude that existing MTD IoT based strategies are weak, and more studies and novel techniques are required to address the specific requirements of IoT systems. 

Overall, MTD techniques have been applied in enterprise systems but their use in an IoT environments remains tentative and requires more research. MTD techniques rely on an important assumption, that the defender has sufficient asset and configuration control that they can change their network configuration without causing any mis-configurations in the process. This is a tall order in large scale and heterogeneous environments and is very likely to be orders of magnitude more difficult for IoT in practice. 

\section{Mediation}
\label{sec:mediation}
Mediation mechanisms allow secure communication while the IoT devices operate in the environment by providing a communication shield between entities.  First introduced by Saltzer and Schroeder in 1975 as a one of eight design principles in the context of engineering secure multi-user operating systems to support confidentiality properties for use in government and military organisations \cite{martinintroduction}, \textit{mediation} is an important concept to ensure control of access, detect anomalous behaviour and address a whole variety of issues including performance and re-configuration. In IoT environments mediation of interactions plays a significantly more important role given the heterogeneity of devices, their resource constraints and the the frequent changes to the system configuration caused by new devices joining, leaving or dynamically connecting to the system. Devices such as IoT Gateways are currently deployed, and the process of interaction mediation goes far beyond the simple traffic filtering commonly encountered in other environments \cite{10.1145/2873587.2873600}.   

Given the inherent distributed nature of IoT environments distribution aspects are important. The traditional design dichotomy between: (i) centralised mediation and (ii) distributed mediation is more present than ever as system designers struggle to resolve the tensions between autonomy, compositionality, control and coordination across a wide array of applications. Centralised mediation is often encountered in systems that see IoT as a separate sub-system that needs to be integrated into the broader Enterprise or Internet architecture. De-centralised solutions emphasise autonomy of control, scaleability and, increasingly, transparency. Naturally, each of the paradigms has its advantages and disadvantages. Centralised mediation is preformed by an entity, usually a resource rich device, called an IoT gateway or an edge device managing the IoT environment and mediating interactions with it. 

The gateway is responsible for performing varied and complex operations, which cannot be distributed easily to the IoT devices without draining their resources, however, it introduces a requirement for the devices to communicate frequently with the gateway, which also can have a high energy cost. Distributed architectures tend to be aimed for devices with more computational and power resources that can coordinate and control their interactions with the environment according to well defined (but possibly dynamically changing) policies. 

\subsection{Centralised mediation (IoT gateway or an edge)}
\label{sec:centralisedmediation}

Several studies discuss a centralised mediation in IoT systems, mostly from the perspective of designing a gateway (IoT edge device) that is not only designed to provide management, interoperability (for example through semantic mediation) and a bridge to allow communication between heterogeneous IoT device, but also to be responsible for securing the IoT system. For example, \citeauthor{6996632} \cite{6996632} introduce a Secure Mediation Gateway (SMGW) for smart home environments that acts as middleware solution to control communication between heterogeneous IoT devices. The SMGW provides secure communications among heterogeneous IoT nodes by means of a federated network comprising two groups, the intraSMGW group and the interSMGWs group. The SMGW acts as a boundary entity between the groups, and allows secure communications among entities within the intraSMGW group and among interSMGW domains. SMGW enables secure remote access to remote devices, and allows remote devices to access a device within the intraSMGW using a secure end-to-end link that adopts a public key and a digital signature encryption schemes. The security services provided by the SMGW include the authentication, authorisation, confidentiality, and integrity for publication and subscription of services in intraSMGW and interSMGW domains. The work in \cite{7412116}, which is based on a previous studies \cite{CASTRUCCI201286} similarly proposes a gateway that provides an abstraction to allow information mediation. Its functionality is further extended to enable heterogeneous Critical Infrastructures (CIs) to predict service failures. 

Gateways are also proposed as a solution to address privacy concerns. \citeauthor{10.1145/2873587.2873600} \cite{10.1145/2873587.2873600} address the issue of privacy and the lack of user control over their local data by proposing a locally-controlled software component called a \textit{privacy mediator} located in the local domain of the sensor devices. The privacy mediator dynamically enforces the privacy policies of the owners of the IoT devices within their local domain, before data is released from the user’s local control. The proposed approach delivers a secure solution at the edge of the cloud, while reducing the privacy burden on the application developers, by involving a small set of trusted parties that provide the mediation code. The framework provides a scaleable solution that can support IoT applications with high data rates.  

Gateways are also seen as a means to address the challenges of heterogeneous communication. \citeauthor{10.1145/3366612.3368122} \cite{10.1145/3366612.3368122} proposes a mediation-based architecture that supports protocol translation to bridge the heterogeneity gap in the IoT network. Although the work does not address issues related to cybersecurity in particular, it tackles the challenge of mediator placement in IoT system. This issue is crucial to solve in order to achieve a timely data exchange between IoT devices. The authors suggest an adaptive placement of mediators through an integer linear programming algorithm that takes into account network resources and IoT device attributes, such as bandwidth, and data size constraints.

\subsection{Distributed mediation and Blockchain approaches}
\label{sec:distmediatin}

Blockchain approaches and Distributed Ledgers have become very popular across a broad spectrum of research studies \cite{al2020survey}. In their original form they are a means of providing non-repudiation without a centralised trusted third party. However, they are now widely applied as fundamental building block of distributed systems, in essence, a distributed database with integrity checking mechanisms. Blockchain and distributed ledger platforms are thus used to build distributed applications for a variety of purposes. The non-repudiation property provided by the blockchain framework has two roles to play in a distributed IoT system. It provides accountability by allowing the posteriori verification of the transactions made. It allows transparency by recording the transactions as well as the policies applicable to those transactions and the policy updates. In a distributed system, policy transparency is an important aspect for coordination. A policy not only constrains the behaviour of components but also allows other system components to have expectations about how other components will behave. In some ways, this is similar to a ``design-by-contract'' approach. 

Several studies found in the literature propose the use of Blockchains as an enabling component of a distributed access control framework. For example, \citeauthor{Ouaddah2016FairAccessAN} \cite{Ouaddah2016FairAccessAN} propose an IoT access control framework named as \textit{FaireAccess} based on Blockchain concepts. The framework adopts the Role Based Access Control (RBAC) model, and defines new types of transactions that are used to grant, get, delegate, and revoke access, and register a new resource with a corresponding address in a private Blockchain. The framework is based on the principles of user-driven, transparency and fairness, without a central authority and aims to provide fine-granularity.  Although the framework shows it can provide transparency of policy updates and their enforcement in a case study, the case-study is really too small to demonstrate the scalability of this approach. 

Similarly, \citeauthor{8386853} \cite{8386853} introduce a framework that investigates the access control in IoT using smart contracts stored in a Blockchain. The work defines  multiple Access Control Contracts (ACCs) that implement both static access right validation based on predefined policies and dynamic access right validation by checking the behaviour of the subject. Again, the case-study used to illustrate the framework is really to small to evaluate the scaleablity of the approach. 

\citeauthor{OUTCHAKOUCHT2017} \cite{OUTCHAKOUCHT2017}, discuss combining Blockchains with artificial intelligence. They suggest the use of distributed access control policies based on one hand on Blockchain technology to enable transparency and distribution, and, on the other hand, on the use of Reinforcement Learning (RL) algorithms to contextualise access control decisions and to achieve dynamic and self-adjusted policies depending on the context that the IoT device operates in. The RL algorithm learns from the written smart contract and takes the best decision for the coming access requests accordingly. However, the work requires sufficiently ``smart'' IoT devices, so that they can run RL algorithms and the blockchain techniques. It is unclear if this framework has been implemented or evaluated in practice. 

The survey in \cite{pervez2018comparative} discusses a number of Directed Acyclic Graph (DAG) based blockchain architectures and their potential in the IoT applications, due to their IoT support capabilities especially in terms if scalability, optimised validation, efficient multi-party involvement.  The discussed techniques are: IOTA, DagCoin, XDAG, Nxt,  Orumesh, Nano, and Byteball. The survey discusses the blockchain architectures for maintaining the ledger and its size, transaction validation,  consensus algorithms, and the scalability. The survey provides recommendations to overcome the drawbacks of blockchain techniques to make them suitable for IoT applications. For example, the survey suggests considering the scalability, flexibility, provenance, low latency, customised configurable  plug and play features considering no transactions fees and costs when designing new blockchain architectures, while supporting  micro transactions, avoiding  centralized monopoly, but providing features to enable IoT devices plugin and portability in various application domains.

\subsection{Continuous monitoring and analysis} 

\label{sec:testing}
Mediation of communication does not only enable controls to be enforced but also to monitor interactions to identify anomalies and identify illegal intrusions through the operation of the system. Continuous monitoring is required to make sure a device continue operating in a safe environment.  Frequently, and increasingly this is done with the help of Artificial Intelligence (AI) techniques, such as machine learning techniques that can learn ``normal'' behaviour and identify anomalous patterns.  

\subsubsection{Intrusion Detection Systems (IDS)} 
\label{sec:ids}

System monitoring typically involves integrating Intrusion Detection Systems (IDS), which are usually combined with IoT Security analytics to monitor the network against malicious system traffic at the network level. There are numerous studies and surveys summarising the work conducted in this area. 

\citeauthor{debar1999towards} \cite{debar1999towards} provides a comprehensive taxonomy for the legacy IDS(s) in networks in general. However, it is important to point out that intrusion and anomaly detection in IoT systems is more challenging than in traditional networks. One reason for this is the diversity of devices and protocols employed, another is the diversity of contexts in which the devices operate, thus leading to large variations in the traffic patterns. Finally, the dynamic evolution of the system e.g., through new devices being added or leaving introduces a further dimension of variation.

\citeauthor{ZARPELAO201725} \cite{ZARPELAO201725} provide a comprehensive survey and taxonomy of IDS(s) in IoT. Generally speaking, IDS(s) are classified as Network based IDS (NIDS) and Host based IDS (HIDS), where NIDSs monitor network traffic connecting one or more network groups or segments and HIDSs are integrated into end devices and monitor system activities. \citeauthor{ZARPELAO201725} further classify IDS(s) according to the IDS detection and placement approaches and the detection methods being applied. IDS detection methods include: 
    \begin{itemize}
         \item Misuse or Signature IDS: This approach requires a database that stores the known threats' signatures integrated with the IDS. The IDS  compares the system and network behaviour against the stored attack signatures, based on which the IDS triggers an alert when the system or network behaviour matches an attack signature stored in the  database. This detection mechanism is evaluated to be accurate in detecting known attacks, it is however, incapable of detecting new threats. Therefore, the signature database is required to get upgraded continuously, which will drain the resources of its host IoT device. 
        \item Anomaly IDS:  following this approach, the IDS is capable of  detecting new threats by comparing the actual behaviour of the system against a profile that depicts its normal behaviour. Given that IoT devices are a special purpose devices and operate according to a predefined profiles issued by their manufacturers, this approach is suitable for IoT environment \cite{ZARPELAO201725}. The IDS triggers an alert whenever the system behaviour deviates from normal and the deviation exceeds a specific threshold. Accordingly, anything that does not match to a normal behaviour is considered an attack leading to a high rate of false positives. Moreover, learning and generating a profile the depicts the system behaviours is not a trivial task.  \citeauthor{ZARPELAO201725} \cite{ZARPELAO201725} discusses a number of anomaly based IDS approaches proposed specifically for IoT environments.
        \item Specification Based: this approach is somehow similar to the anomaly based detection approach, as the IDS identifies deviations from normal behaviour by comparing the system behaviour against a manually defined system specification.  Specification determine the set of rules and thresholds that define the expected behaviour of a network and system components. This approach only flags behaviour that is outside the specification and thus provides lower false positive rates than anomaly-based detection but is less specific. Whilst this detection method can start working immediately as it does not need a training phase, maintaining manually defined specification of device behaviour and expected physical invariants is challenging, difficult and error prone. 
         \item Hybrid detection IDS: IDS hybrid detection approaches harness the advantages of the discussed signature-based detection, specification-based and anomaly-based detection and attempt to combine them in one IDS. 
    \end{itemize}

IDS placement approaches include:
    \begin{itemize}
        \item Centralised placement: in this approach, an IDS is installed in one device, usually a dedicated resource rich device, such as a gateway or a router. This allows to monitoring the network traffic between the IoT devices and the Internet. This approach is less effective at capturing an analysing device-to-device interactions and the central may not detect attacks that compromise part of the network \cite{ZARPELAO201725}.
        \item Distributed placement: by adopting this approach, the IDS is installed in each single device in the network. It is also possible for each device in the system to monitor its neighbours, those devices are called \textit{watchdogs} and are commonly used in data-spoofing detection approaches. IDS(s) in this type of placement have to be lightweight to minimise the resource consumption on the end devices. 
        \item Hybrid placement: in this approach, the network is partitioned into clusters, and only cluster heads of each cluster (which are resource rich nodes) are responsible of monitoring their associated clusters. This aims to combines the advantages of both centralised and distributed approaches. 
    \end{itemize}

\subsubsection{Honeypots}
\label{sec:honeypots}

Applying IDS should be combined with applying honeypots, to form what is called the ``Intrusion Detection Honeypot''. A honeypot is a trap used to attract attackers to exploit fake vulnerabilities in order to gain access to a fake target. Not only does this provides early warnings once a threat is detected, but also helps in collecting, analysing and identifying information about attackers and their activities. Spitzner in \cite{Endpoint} best defined a honeypot as "a resource whose value lies in being probed, attacked or compromised". There are different types of honeypots, and they can be categorised in different ways; a common approach is to consider their level of interaction with the attacker. High-interaction honeypots allow full access to a ``real'' system i.e. one that reproduces with a high fidelity an actual real system's operation and data. Low and medium interaction honeypots reproduce the system's behaviour and data with lower fidelity but as a result are easier to set-up and maintain \cite{9520645, Vetterl2020HoneypotsIT}. This makes them very convenient to use for detecting and gathering information about large scale attacks. 

Honeynets comprise multiple honeypots deployed within the same system. \citeauthor{9520645} \cite{9520645} provide a comprehensive review of the literature related to honeypots and honeynets for IoT environments, Industrial Internet of Things (IIoT), and Cyber-Physical Systems (CPS) over the period 2002-2020. The survey introduces a taxonomy of honeypots and honeynets based on their: purpose, role, level of interaction, scaleability, resource level, availability of source code and target (IoT, IIoT, or CPS application). \citeauthor{Vetterl2020HoneypotsIT} \cite{Vetterl2020HoneypotsIT} also investigate low and medium interaction honeypots proposed in literature but also design a high interaction honeypot for IoT and Customer-premises equipment (CPE) called \textit{Honware}. \textit{Honware} can be implemented and deployed rapidly as it extracts firmware images automatically to emulate the device's behaviour in a virtual environment. Its evaluation revealed that it is able to detect known and unknown attacks effectively. \textit{IoTPOT} is another newly developed high-interaction honeypot system designed to mimic IoT devices, and is used to analyse the behaviour of botnets \cite{191952}. \citeauthor{Vetterl2020HoneypotsIT} \cite{Vetterl2020HoneypotsIT} also reviews other honeypots proposed in literature for IoT systems, including \textit{Conpot}, which emulates industrial control systems \textit{HoneyPhy}, which simulates attached physical devices. In contrast to \textit{Honware}, \textit{IoTPOT}, \textit{Conpot} and \textit{HoneyPhy} do not use actual firmware to emulate the real IoT devices and only emulate specific application and network layers. Thus their behaviour is less similar to that of actual IoT devices. 

In contrast to IDS, a honeypot does not have any legitimate traffic, as there is no reason for legitimate traffic to connect to the honeypot. This means that malicious activities are the only ones honeypots can detect, and any activity with the honeypot is likely to be a threat or an intrusion attempt, leading to a low false positive rate. This stands in contrast to the traditional IDS which can generate a high level of false alerts. This also means that with using honeypots, the real asset may not be touched at all by the attackers leading to a more secure system. Additionally, honeypots provides a mean for distracting the attacker from going forward towards the actual target by spending more time in trying to exploit the wrong system. Not only this would exhaust the attacker's resources, but  would give more time for the security experts and system administrators to identify attacker's intent and techniques and to isolate the actual target. However, there are no guarantees that attackers will ``fall for it''

If implemented or deployed inadequately, honeypots may actually increase the risk to the system as they can attract attackers. For this reason and to combine their efficacy at isolating attacks with the need to reduce the false positive rates of IDS(s), honeypots should always be introduced in conjunction with rather than as an alternative to other security approaches such as IDS and firewalls. 

Note also that low interaction honeypots are easily detected by attackers. For example, \citeauthor{Vetterl2020HoneypotsIT} \cite{Vetterl2020HoneypotsIT} report that their study was able to fingerprint a large internet-scale low and medium interaction honeypots, which use off-the-shelf libraries for protocols implementation. The study also found that a large number of honeypots are out of date, and their operators use a standardised deployment scripts, which facilitates detecting them. For these reasons, the study recommends the development of a new generation of honeypots, which focuses on the lower levels of the network stack. Another study \cite{Pauna2019OnTR}, also reports that modern attacks use multiple interactions with the target prior to the attack step itself. Thus, to be credible and not detected, honeypots require a higher degree of fidelity.

\subsection{Artificial Intelligence (AI) techniques for monitoring and detection}
\label{sec:AI}

Nowadays, combining AI with security methods is a favoured approach to dynamically and accurately secure systems against threats. This is due to rapid evolution of attacks in the IoT environment in particular, and the need to protect these systems intelligently and in real-time. For example, it is increasingly popular to combine IDS(s) with machine learning (ML) approaches, that help updating the system autonomously in order to detect and prevent new type of attacks \cite{kumar2021uids}. 

\citeauthor{kuzlu2021role} \cite{kuzlu2021role} discuss the role of incorporating AI in IoT cybersecurity, and the use of machine learning methods such as decision trees, K-nearest neighbours, support vector machines, and neural networks to detect attacks in IoT environments. Similarly, \citeauthor{chaabouni2019network} \cite{chaabouni2019network} discuss NIDS techniques employing ML for IoT environments and provide a comprehensive review of NIDS(s) that integrates different aspects of learning techniques for IoT. \citeauthor{kumar2021uids} \cite{kumar2021uids} propose a \textit{unified intrusion detection system for IoT environment (UIDS)} to defend the network from attacks also based on ML approaches. The proposed IDS model acts as the watchdog in the IoT based system to prevent the system from internal and external malicious attacks. \citeauthor{meidan2018n} \cite{meidan2018n} present a NIDS method that uses deep auto-encoders to detect anomalies in the IoT traffic. The approach is  evaluated  empirically on different commercial IoT devices considering IoT botnets such as \textit{Mirai} and \textit{BASHLIT}. \citeauthor{pacheco2019security} \cite{pacheco2019security} propose an IoT threat framework with a neural network model that is able to detect potential attacks against each layer of the framework. They create a reference model for end nodes automatically and perform behaviour analysis at run-time. The authors claim that the framework recognises both known and unknown threats with high detection rate and low false positive alarms.

Honeypots have been evolved to include AI and ML techniques to interact with the attacker and decide the actions to be taken, instead of relying on human pre-programmed traps. These are referred to as \textit{``self-Adaptive'' honeypots}.  For example, \citeauthor{Pauna2019OnTR} \cite{Pauna2019OnTR} propose a self-adaptive IoT honeypot system that interacts with the attackers based on a set of actions triggered by a reinforcement learning algorithm.  They test two behaviours that partially map the Mirai botnet.

Despite their benefits, AI techniques usually requires complex operations and therefore requires resource rich hosts, which limits their usage in the endpoints of the network. Moreover, AI techniques vulnerable to attacks, and can even ``weaponise'' the AI itself against the system. Methods that exploit AI to attack IoT systems are discussed in \cite{kuzlu2021role}.

\section{Device(s) isolation and system segmentation}\label{sec:segmentation}

When an attack is detected, in order to eliminate it from spreading further into the rest of the IoT system and other connected systems, then the infected devices and all (or part) of the devices reachable from the compromised nodes should be isolated from the network. Putting devices into quarantine thus seeks to eliminate the threat from infecting the rest of the network and further the damage, whilst allowing the rest of the network to continue working after having being partially compromised \cite{10.1145/3411498.3419967}.

This is, however, a delicate approach, as inter-dependencies between devices within the IoT system and to other parts of the system can aggravate the effect of isolation.  Identifying the dependencies between IoT system components and other systems, and modelling and analysing the system risk, are required to evaluate the level of threat across the entire network, and prioritise which reachable devices need to be isolated and which are not. This helps to perform device isolation without unduly degrading the performance of the system as a whole. Segmenting and segregating the network is a always a good practice to mitigate risks when designing a network, but one has to make sure that the benefit gained from isolating the infected devices exceeds the losses \cite{soikkeli2019efficient}. Therefore, the first design requirement to be considered when constructing the system, and before segmenting the network is to identify and document IoT dependencies.

\subsection{ Identifying and documenting IoT dependencies}
\label{sec:dependencies}
IoT attacks typically have cascading effects on the system, due to the inter-dependencies between the system's devices and components. In many cases, IoT systems are basically systems of systems or systems connected to critical infrastructure (CI). Thus, failures in one system leads to disruptions in other systems, which can unpredictably proceed further to causing severe unexpected or undesirable consequences \cite{stellios2018survey, 9174628} . 

Despite their critically, especially in CI or with systems connected to CI, systems inter-dependency and possible failure mechanisms are not well understood and have only recently begun to receive attention in the IoT community. \citeauthor{little2003toward} \cite{little2003toward} identifies three classes of infrastructure failure effects due to inter-dependencies between systems: (i) The cascading effect, when a disruption in one system causes a disruption in another, (ii) the escalating effect, when a disruption in one system exacerbates an independent disruption of another, and (iii) the Common cause effect, when a disruption of two or more systems occurs at the same time because of a common cause. \citeauthor{9174628} \cite{9174628} reviews the literature on cascading failures in IoT systems, their major causes (one of which is the cyber-attacks), their mathematical models and simulations, and mitigation techniques in IoT systems that take into account cascading failures from a reliability and resilience perspective. They consider how IoT system inter-dependencies can affect the resilience of the system. The article also sheds light on the differences and relation between reliability and resilience and states that reliability is the end goal of system design while resilience is the way in which the end goal can be achieved. 

\citeauthor{10.1145/3302505.3310082} \cite{10.1145/3302505.3310082} discuss the importance of considering dependencies on network connectivity and services in IoT systems, and explore the service and infrastructure dependencies from a cloud service and geographical perspective in smart IoT environments over a longer time period. In particular, the work studies the robustness of IoT device operation when connectivity is disrupted, and reveals that many IoT devices make use of services distributed across the globe, and are thus dependent on the global network infrastructure, even when they performing purely local actions. Consequently, some devices do not operate properly without network connectivity, even if their behaviour requires only local information. The study also reveals that IoT devices behave differently when connectivity is recovered after some disruption. It recommends that dependencies in IoT systems need to be better considered, and documented in their design and implementation processes respectively, to obtain a clearer picture about the impact of IoT device deployment on infrastructure resilience.

\citeauthor{stergiopoulos2020automatic} \cite{stergiopoulos2020automatic} model the inter-dependencies between assets and devices in IoT environments, and develop a framework that presents those connections in a company’s business processes. The framework discusses automation in security risk analysis and restructuring IoT systems, by leveraging dependency risk graphs, graph minimum spanning trees, and network centrality metrics to model the networks' dependencies. The framework was tested on a real-world company, and proved its ability to automatically identify critical components and dependency structural risks, prioritise assets based on their impact on business processes and propose network topologies with the optimum number of asset sub-nets, while preserving business operations. Similarly, the studies reported in \cite{eusgeld2011system, alcaraz2015critical}, discuss inter-dependencies in systems with underlying critical infrastructures, such as industrial control systems (ICS), SCADA (Supervisory Control and Data Acquisition), and explores vulnerabilities, threats, and challenges related to critical infrastructure protections including security management, self-healing, modelling and simulation, forensics and learning. \citeauthor{luiijf2021analysis} \cite{luiijf2021analysis} discuss Critical Information Infrastructure (CII) disruptions, dependencies, and cascading effects based on empirical data. \citeauthor{cardellini2007agent} \cite{cardellini2007agent} attempt to model inter-dependencies in complex systems and critical infrastructures through agents and simulation, using UML as their modelling language.

\subsection{Network Segmentation}

Identifying inter-dependencies helps to identify critical system devices, prioritise assets and document inter-connections between those components and other systems and critical infrastructures. Such information is critical to determine the impact of isolating a device or a group of devices when threats occur, or in order to avoid cascading disruptions to the system. Isolating groups of devices in a network, in its broader context, is known as \textit{network segmentation}. Generally speaking, segmentation is partitioning the network, system or environment into smaller groups or networks (segments), sometimes even down to the host itself. 

Network segmentation is a well known practice commonly applied in legacy networks before the emergence of IoT, and can help contain security breaches to a single (or a few) infected segments, thus preventing the threat from spreading further into the other parts of the network. The Information Assurance Directorate (IAD) identifies two approaches for network segmentation \cite{7849908}: 

\begin{enumerate}
    \item Segregate Networks and Functions (SNF), which refers to partitioning the network into groups or segments based on their functionality or intended services, and limiting the communications between the segments. This hinders the attacker, who managed to gain access into the network, from causing harm by acquiring further access into the rest of the network. 
    \item Limit Workstation-to-Workstation Communication (LWC), which refers to controlling communications on a more granular level inside a segment, between segments and between segments and the Wide Area Network (WAN). LWC grants communication privileges only when necessary by enforcing the least privilege principle.
\end{enumerate}

Legacy practical approaches to network segmentation include both physical and logical methods. The physical approach uses multiple firewalls, for example, to separate financial applications from medical devices and IT applications within a hospital. This approach remains expensive and requires complex separations with thousands of firewall rules to segment internal networks. Logical segmentation on the other hand, is conducted with VLANs, whose configuration is equally labour intensive. Network segmentation is traditionally conducted as a proactive approach once the network is constructed. For the enterprise networks, a number of studies have investigated means of automating network segmentation. \cite{10.5555/3106388.3106405,10.1145/3205651.3208287, doi:10.1177/1548512916662924, 7849908, Wagner2017AND}. 

Due to the unpredictable and highly dynamic nature of IoT environments, IoT network segmentation offers a promising approach to defending them by adapting the system in real time and implementing reactive approaches through \textit{dynamic segmentation}. However, very few work studies tackle dynamic network segmentation, e.g., \cite{Wagner2017AND}, and even then are oriented to Enterprise systems. The approach proposed in \cite{Wagner2017AND} generates the segmentation architecture dynamically when threat level changes, based on a nature-inspired process that investigates different solutions using Simulated Annealing (SA). In their proposed technique, \cite{Wagner2017AND} adopts a continuous time Markov chain model (CTM) to evaluate the risk in a network environment, and generates architectures adaptively by adding, removing and altering the existing services in order to combine or split enclaves containing the devices. The technique segments the network according to a combined risk measure that takes into account the segmentation cost as well as the security risks. Risk minimisation is the primary goal to the optimisation function. However, the work only considers attacks originating from the Internet, whilst in the context of IoT, new (compromised or infecting) devices could dynamically join and leave the network enclaves. \citeauthor{GEORGE2019101068} \cite{GEORGE2019101068} proposes a methodology for risk assessment in IoT environments that considers multiple attackers and multiple targets. Their mitigation technique recommends removing the edge device and removing the minimum number of vulnerabilities necessary to secure the target. The work adopts a greedy algorithm that finds and removes the dominant attack paths between the attacker to the target. However, the proposed risk assessment and mitigation approaches do not tackle the dynamic changes of threats in the IoT environment.

As mentioned earlier, network segmentation has not been applied broadly to IoT networks. There are a number of reasons that discouraged the implementation rate of network segmentation in IoT systems. One of which is the administrative complexity of applying segmentation by the average user, specially in personal settings, such as personal or home networks. Another, is that network segmentation approaches require significant manual and expert effort. Dynamic segmentation techniques such as those mentioned earlier cannot easily scale to large network and they cannot cope with the dynamicity of the IoT environments. IoT devices frequently exchange information with each others and with the cloud, breaking the perimeters usually secured by residential firewalls or gateways \cite{wasicek2020future}. 

More importantly, IoT devices are always active in a highly connected environment, where they are hardly managed or patched against the latest security updates, which all lead  to an increased system attack surface. If an attacker manages to pass the residential firewall, which in many cases is the only firewall in the network, and succeeds in exploiting one entry point in the system device, the attacker then perform lateral movements to connect and infect other devices in the system without restriction. Consequently, the whole system including other devices connected to the same network are at risk. Lowering the manual effort to perform segmentation is key, but research approaches on dynamic network segmentation still rely on complex and highly extensive network simulations that do not scale well. 

More recent approaches such as \textit{Micro-segmentation} (see next section) use the capabilities of SDN coupled with edge cloud to implement segmentation for individual devices and devices' workload. 
  
\subsection{Host segmentation (micro-segmentation)}
\label{sec:microseg}

Classical network segmentation approaches based on firewalls, VLANs etc. have difficulty coping with the dynamic nature of IoT environments where devices can frequently join and leave the the network dynamically. At the same time the network infrastructure is becoming more programmable e.g., through Software Defined Networking (SDN) and Network Function Virtualisation (NFV), while edge networks such as home networks are increasingly supported by an \textit{Edge Cloud Network} such as residential gateways in 5G networks. By adopting micro-segmentation, devices in IoT systems can be individually characterised and isolated to prevent a weak device from serving as an entry point to the entire network. While both network segmentation and micro-segmentation control the flow of traffic between network segments and application components based on security rules, micro-segmentation works at a finer level of granularity segmenting the individual host workloads throughout the life-cycle of the device from it first joins the network until it is disconnected. Therefore, micro-segmentation is often referred to as "host-based segmentation" rather than "network segmentation".

Micro-segmentation establishes a network inventory of all the devices. When a device joins the network it is fingerprinted and scanned for vulnerabilities and registered in a \textit{network inventory} virtual network function. By default, the wireless traffic is configured to run in client isolation mode and a \textit{micro-segmenter} reprograms the smart gateway via a protocol such as OpenFlow. \citeauthor{wasicek2020future} \cite{wasicek2020future} highlights the importance of applying micro-segmentation in 5G enabled-smart home IoT environments, so as to reduce the attack surface and protect such environments from internal compromise involving lateral movements. The system realised in the context of home networks is described in \cite{253364}. The authors implemented micro-segmentation in an emulated network topology, which included both IoT and non-IoT devices belonging to six functional groups (energy, management, controller/hubs, cameras, appliances, health monitors and non-IoT). The adopted micro-segmentation approach identified and quarantined infected devices from accessing the LAN and WAN, whilst the non-malicious devices were automatically classified based on functionality and assigned to confined network micro-segments accordingly. 
The work found that micro-segmentation reduces the attack surface exposed to a webcam infected with Mirai botnet by 65.85\% compared to the baseline configuration, which was at the expense of preventing 2.16\% valid network flows between devices. This deviation resulted from flows that would cross the functional micro-segments. 

\citeauthor{finproceedings} \cite{finproceedings} also discuss the concept of micro-segmentation and its integration in 5G, and compare it with the concept of network slicing. The authors implemented a virtual micro-segmentation, and a network slicing approach in an experimental test bed, which includes personal IoT health applications, and discuss the adminsitration of the micro-segments. This work also discusses the different trust models that can be adopted, such as the now popular \textit{Zero Trust} model. 
Inherently, \textit{Zero Trust} adopts a least privilege approach, and is best implemented in IoT applications with critical services, such as e-health. Different authentication and verification levels can be adopted in micro-segmented networks, and the authors suggest that micro-segments could have different security levels depending on the application and the service provided. For example, micro-segments for critical applications such as e-health could operate with different security requirements compared to non critical IoT micro-segments.

\citeauthor{10.1007/978-3-030-81111-2_11} \cite{10.1007/978-3-030-81111-2_11} discuss a standard proposed to implement micro-segmentation in IoT systems, which is the \textit{Manufacturer Usage Description (MUD)}, and which was proposed by the Internet Engineering Task Force (IETF). The MUD standard enables manufacturers to specify the intended use of their devices, thereby allowing them to be placed in the appropriate micro-segments. The standard facilities an automated protection and micro-segmentation for IoT systems by using dynamic access control lists. Those lists contain rules for every authorised access for each device in the IoT network. The work discusses an interactive system called MUD-Visualiser that visualises the files containing these access control rules, in order to increase usability.

\textit{Network Segmentation} and \textit{Micro-Segmentation} are complementary approaches as they operate at different levels. Both rely on the principle of mediating network access and inter-connectivity and increasingly rely on the dynamic programmability brought about by techniques such as Software Defined Networking (SDN) and Network Function Virtualisation (NFV). 

 \subsection{Software Defined Networks (SDN)} 
 \label{sec:SDN}
The application of SDNs techniques in IoT scenarios has attracted a significant interest in recent years. \citeauthor{inbook} \cite{inbook}, present a light-weight policy-based approach designed as part of the \textit{SerIoT} Research and Innovation Project, funded by the European Commission. The proposed system called \textit{Autopolicy}, proposes an architecture for enforcing different traffic profiles according to the intended communications of an IoT system with other devices or systems in order to mitigate potential security risks. The \textit{Autopolicy} system integrates a distributed machine learning approach based on deep learning (DL) and graph networks for risk monitoring, which analyses the network traffic in real-time. This SDN and AI integration realises a \textit{Self-Awareness} that can achieve secure and QoS-based routing of traffic flows. Such efforts are aligned with the recent MUD standard, mentioned earlier. \citeauthor{garcia2019enforcing} \cite{garcia2019enforcing} propose an architecture for managing, obtaining and enforcing MUD restrictions on SDN switches. The work analyses the applicability and advantages of using MUD in industrial environments, and provides a comprehensive performance evaluation of the processes required. \citeauthor{zarca2019security} \cite{zarca2019security} design an architecture that captures the security and privacy requirements in cyber-physical systems and IoT-CIs and make autonomous security decisions and re-configures the the network using SDN and NFV. The reconfiguration in response to IoT threats is done through special purpose IoT agents, SDN and IoT controllers as well as NFV equipment, which enforce security countermeasures and dynamically adapt the system according to the context analysed by the integrated monitoring tools. This architecture has been implemented and evaluated in smart buildings, as part of the \textit{ANASTACIA} H2020 EU research project. For more information about SDN and NFV, I refer the reader to \cite{alonso2019survey}, which discusses extensively the characteristics of SDN and NFV and their potential to secure IoT environments.

Whilst SDN and Micro-segmentation leverage mediation and programability as the main characteristics to segregate and ensure security in dynamic IoT environments, this comes at a significant cost in complexity and management of this programmable network environment. Two aspects remain, in my view, as yet less well understood. Firstly, is it possible to adapt such dynamic configurations to the actual network and device usage in large environments? Device function may not fit neatly into segmented boundaries and may also change over time. Secondly, what are the vulnerabilities and potential for new threats introduced by such techniques? Their potential for being abused is not investigated as thoroughly as their use in support of security functions.    

\section{System availability, Redundancy and Resilience} 
\label{sec:availability}
As we increasingly rely on IoT for sensing the physical environment and adapting to it, we rely on the \textit{availability} of the system even when it is subject to threats and compromise. Ensuring the \textit{resilience} of the system entails minimising its loss of function over time and thus maximising its availability including when under threat, whilst at the same time ensuring the integrity of the functions delivered. Such resilience must be ensured for the system itself as well as for the data it delivers and on which decisions are based. To ensure resilience, especially in an IoT system where devices can be trusted to various extents, requires redundancy to absorb failures and compromises. This is particularity important if the system incorporate sense sensitive data.   

\subsection{Redundancy vs. diversity} 
\label{sec:diversity}

Redundancy is commonly applied in \textit{fault-tolerant systems}. Applying redundancy means deploying additional system components identical to other systems' components. This method guarantees the system availability even if fault occurs in the original component. However, having identical redundant components is not helpful if the system is exposed to malicious interventions as identical components will also share the same vulnerabilities. If an attacker manages to exploit a vulnerability in one component, they can easily exploit the same vulnerability in its identical copies for negligible additional cost. Redundancy must therefore be coupled with ensuring \textit{diversity} in the redundant resources, seeking to minimise the shared vulnerabilities. Note that a different form of diversity, in the network configuration rather than in the component provision, is ensured by \textit{MTD} approaches discussed earlier in Section \ref{subsec:mtd}. 

Redundancy, diversity and other key enablers for system resilience are discussed in \cite{dobson2019self}, which focuses on the design principles in the area of self-organisation and resilience of networked systems. The authors have also previously published a survey, which discusses redundancy, besides diversity and connectivity as key enablers to achieve network resilience in terms of fault-tolerance, survivability, and performability \cite{sterbenz2014redundancy}. The survey highlights that diversity is essential to provide survivability in addition to using redundancy for fault tolerance and connectivity which is recommended for disruption tolerance. Diversity provides alternatives, which can run simultaneously or as needed, so that even when a particular alternative is compromised, other alternatives can provide some degree of functionality. The survey also introduces a cross layering model that leverages these techniques to achieve a resilient network system. Similarly, \citeauthor{8539114} \cite{8539114} propose a framework that leverages redundancy, diversity, and hardening techniques to improve the security and resilience in IIoT. The framework introduces a threat model, which quantifies security risks and the impact of attacks in order to prioritise security investments in redundancy, diversity, and hardening. The prioritisation is formulated as an optimisation problem, which proves to be an NP-hard problem. The work evaluated the applicability of the framework in a water-distribution system, and their results revealed that integrating redundancy, diversity, and hardening helps reduced security risk whilst maintaining the same cost. \citeauthor{venkatakrishnan2016using} \cite{venkatakrishnan2016using} also investigate the use of redundancy with diversity, but focus on detecting ``difficult-to-detect'' attacks on web servers deployed as part of IoT applications. They conclude that, although the redundancy-based detection technique maybe costly, it is worth considering in IoT environment, especially when the system needs to operate for long periods of time without direct human intervention. The work also recommends masking design differences when adopting a redundancy-based detection approach. 

Despite the benefits of having additional diverse or redundant resources in improving the resilience of the IoT system, applying those methods can be costly, especially in applications where IoT networks are connected to critical infrastructures. Applying diversity also significantly increases system complexity \cite{linkov2019fundamental}, as it introduces additional heterogeneity and makes it harder to manage the system in a uniform way. In the following subsections, I discuss alternative methods to account for data loss, that have potential in-providing a form of data backup without increasing the cost, for example by leveraging the  heterogeneity of IoT data and applications. 

\subsection{Exploiting data correlation}
\label{sec:dataredundancy}
Data resilience seeks to leverage data redundancy in order to ensure data integrity and availability. Redundancy in this context can manifest through monitoring the same measurements through different sensors (e.g., in Wireless Sensor Networks) or through correlations in the measurements made. Such correlations can be in time, in space, or across the different attributes measured by the sensors. For example, in Wireless Body Area Networks (WBAN), both the ECG and hemodynamic signals, such as blood pressure, have information mutually correlated due to the physiological inter-relation of the mechanical and electrical functions of the heart \cite{salayma2017wireless}. Such correlations enable the detection of measurement manipulation, an area known under slightly different terms \textit{data spoofing}, \textit{malicious data injection} or \textit{false data injection} in different research communities. For a survey of such approaches in Wireless Sensor Networks see \cite{lupuillianosurvey}. Whilst a number of techniques have been developed for correlations in time series and further techniques have been developed for exploiting spatial correlations \cite{Illiano2017DontFM}, the topic remains less investigated in IoT applications where few sensors can be deployed and they measure heterogeneous information. Information fusion, and the ability to exploit at the same time temporal, spatial and attribute correlations remains to be investigated further.   

Data correlations, and thus information redundancy, is often exploited to detect malicious manipulations and thus sensor compromise. This can be achieved by detecting anomalies in the expected correlations between measurements that data spoofing introduces \cite{Illiano2017DontFM}, or more commonly in cyber-physical systems by building estimators from measured values and knowledge of the physical phenomena and comparing these estimators with the measured values of other sensors \cite{10.1145/3203245}. Less common are information fusion approaches that seek to ``repair'' the compromised data to preserve data availability. 

Exploiting data redundancy has a cost, more specifically that of measuring, transmitting and aggregating correlated, and thus redundant information. There is therefore a tension between the resilience requirement in IoT applications and the requirement to minimise function and energy costs. How to characterise this trade-off in a generic way is something that needs to be further investigated. Another consideration arises from the necessary contextualisation of the measurements and the information acquired. The correlations between the measurements vary with the physical phenomena being measured and with the deployment context of the WSN. For example, in healthcare applications the correlations vary according to the activities undertaken by the patient. In fixed, wireless sensor networks, e.g., such as those monitoring temperature or volcanic eruptions, the correlations vary when an eruption happens \cite{Illiano2017DontFM}. Establishing a base-line in a network guaranteed to be uncompromised, remains a significant challenge.   

\subsection{Exploiting data overhearing}
\label{sec:overhearing}
Overhearing is a phenomenon usually occurring in wireless sensor networks (WSN), where each active, or idle sensor node overhears data not designated to itself but sent by neighbouring nodes. Overhearing is, in essence, a side effect of the broadcast nature of the wireless medium. Although this phenomenon has often been considered as drawback in WSN, as it causes extra energy consumption and therefore leads to fast energy depletion, data overhearing can be exploited to overcome data loss as it can provide a form of redundancy. This allows two or more network devices to help each other in transmitting the information to a common destination. Such approaches are commonly referred to as \textit{cooperative network}. This cooperation (or collaboration) between network nodes can be useful especially in critical IoT applications where data has to be delivered reliably and with high priority. If an attack causes data loss in such applications, overhearing can be exploited to account for the lost data. Despite its potential to ensure the reliability and resilience of data transmission this area remains under investigated, especially in the context of cyber attacks. 

\citeauthor{9469421} \cite{9469421} investigate the potential of exploiting redundant information using collaborative methods to mitigate data loss in LoRa network. In this work, a centralised gateway encodes data frames sent from nodes, which combine their neighbours overheard data into their own transmissions. Two collaborative approaches are proposed: (i) Neighbour Data Re-transmission (NDR) in which a node appends its most recently overheard frame to its subsequent frame to be transmitted, and the (ii) Combined Data Re-transmissions (CDR), in which a node combines its most recently overheard frame and its subsequent frame using a bitwise exclusive-OR (XOR) i.e it applies Network Coding (NC). The work evaluated its potential to mitigate data loss through both simulations and empirical studies. \citeauthor{8412519} \cite{8412519} propose \textit{LeapFrog Collaboration (LFC)}, a communication protocol that exploits overhearing between network nodes and relays, and uses packet redundancy to account for packet loss in wireless industrial networks. The work duplicates the data flow onto two paths allowing the nodes on one path to overhear the data transmitted along the other path. LFC provides reliable communication over RPL based networks, as each transmitted packet has multiple chances to be successfully delivered. Simulation results revealed that LFC achieves high network reliability e.g., above 99\%.

Overhearing is also exploited to detect malicious data injections. In essence, each node overhears the transmissions of their neighbours and compares the measurements reported with the ones they have measured themselves. Trust-based algorithms have then been proposed to allow sensor nodes to vote on the trustworthiness of their neighbours based on the results of this comparison. Such approaches have been often proposed in the literature but suffer from an intrinsic limitation: a compromised node can lie about the measurements or about the trustworthiness of its neighbours or both. Distinguishing attacks on the basis of data analysis alone becomes then very difficult. 

Overhearing also introduces a tension between the data availability and confidentiality objectives. Overhearing allows data to be replicated and re-transmitted but this offers attackers more opportunities to access the data through sensor or network compromise. Finally, overhearing introduces additional energy costs both to listen to the neighbours' transmissions and to re-transmit the data. 

\subsection{Network Coding (NC)}
\label{sec:coding}
Network Coding is commonly covered in information theory, and has many practical applications in networking systems. It is often used to increase the throughput and robustness of a network by using diverse paths for different packet combinations \cite{ahlswede2000network}. Using NC, the nodes recombine input packets into one or more output packets. This allows intermediate nodes to transmit packets that are linear combinations of the previously received packets instead of mere forwarding packet by packet to the same destination. Hence, NC provides a form of data redundancy that not only can be used to mitigate data loss in lossy links, but also to recover lost data whilst reducing the energy costs required by the transmission. In particular, it avoids unnecessary packets re-transmission and acknowledgements. As a result, the application of network coding in Internet of things can also contribute to ``green IoT networking’', whilst providing a degree of resilience \cite{li2017towards}. 

\citeauthor{inbookt} \cite{inbookt} proposes a network coding-based protocol (NCBP) to detect and recover lost packets and correct errors in an IoT networks. NCBP is based on a random network coding (RNC) scheme that allows nodes to generate linear combination of input packets into coded packets separately over a finite field. The work compares the performance of NCBP with that of legacy algorithms such Forward Error Correction (FEC) and Automatic Repeat Request (ARQ). Their results show that NCBP can effectively recover lost data, increases throughput with less re-transmissions and minimising delay, bandwidth and energy consumption, thus emphasising its suitability in IoT environments. \citeauthor{peralta2018network} \cite{peralta2018network} also discuss NC-based techniques and their potential in next-generation IIoT architectures. The study emphasises the benefits of applying NC in IIoT, and investigates the suitability of NC to improve network performance in environments with unstable channel conditions. It also emphasises the suitability of NC-based techniques in dynamic topologies and constrained environments.  

\citeauthor{8664097} \cite{8664097} investigates the use of NC for data recovery in IoT networks focusing on the detection and storage requirements. The work proposes an eavesdropping prevention technique that combines NC and Recurrent Neural Network (RNN), and models an optimisation problem for minimising device storage subject to meeting a set of security requirements. The work proposes two allocation algorithms for IoT device storage failure prediction; the Failure-Aware Greedy Allocation (FAGA) and the Failure-and-Load-Aware Greedy Allocation (FLAGA). It evaluates the performance of the proposed techniques using a real dataset, and shows that the proposed technique can meet strong security requirements.

\citeauthor{zverev2019systematic}  \cite{zverev2019systematic} focused on NC approaches with overlapping generations and studied the the possibility of modifying the number of redundancies per generation. Their simulation results show that increasing the overlap is more efficient than having a larger number of redundant encodes per generation, but at the expense of a larger the overhead. Therefore, the work recommends carefully considering the trade-off between network efficiency and delay.

\citeauthor{akilandeswary2020next} \cite{akilandeswary2020next} review the stat-of the-art related to Random Linear Network Coding (RLNC) approaches in IoT environments. In RLNC, the sender combines the incoming packets with the help of randomly generated coefficients, while the receiver recovers the lost packets, with the help of an encoding vector, which helps the receiver know the coding coefficients used by the sender during the encoding process. The study has implemented four different variants of RLNC: (i) Multicast sender-receiver (ii) Encode Decode using Random Coefficients (iii) Recoding (iv) Encode/Decode on the fly. The authors point out the need to develop a generic and adaptive RLNC system, that can change according to the data transmission requirements.

\citeauthor{peralta2019homomorphic} \cite{peralta2019homomorphic} investigate the integration of NC with Homomorphic Encryption (HE) to enhance privacy and confidentiality whilst improving the communications and data distribution across the entire IoT architecture. The study reviews the related literature that address the role of both technologies in IoT systems, and discusses the benefits and challenges of integrating those technologies in IoT environments. One disadvantage for applying NC is that complex NC schemes can lead to additional delays, which do not meet the requirements in real-time applications. 

The main drawback of NC-based techniques is that they are based on trust among nodes. It is easy for any malicious node to join the network and act as an intermediate node that could forge encoded packets. The receiver may not be able to detect attacks, and hence may not be able to recover the original data packet correctly under pollution attacks, but will attempt to reconstruct from the wrong data. What makes it even more difficult to distinguish the valid encoded packets from malicious ones, is that the packets received by the receiver are combined with several other packets originating from multiple sources. In an attempt to address this issue, \citeauthor{Lee2020AttackDU} \cite{Lee2020AttackDU} propose a method to detect compromised packets among the packets received at the receiver in an IoT environment that adopts NC for data recovery. Their detection method enables the receiver to identify the valid packet amongst the ``look-like-valid'' packets without requiring re-transmissions. The proposed scheme shows that a receiver can recover a valid packet with a high probability when there is sufficient redundancy.

\section{Conclusions}
\label{sec:conclusions}
The security and resilience of IoT environments remains a complex topic that spans across the entirety of their life-cycle from design and realisation to their deployment operation and decommissioning. The challenges are being addressed in the design of individual devices and in the design and operation of deployments. Like in the case of enterprise or more traditional computing environments, new techniques are being developed to make devices more trustworthy and new techniques are being developed to make systems more resilient and trustworthy by detecting, mitigating and responding to threats at run-time. 

However, the IoT presents unique challenges that do not encumber the development of solutions in traditional systems to the same extent. The heterogeneity of devices and contexts of use makes it difficult to develop and apply common standards e.g. for design, deployment or certification and to develop general solutions. The increased dynamicity of the environment makes statically planned architectures and frameworks more difficult to employ. Dynamic adaptation and continuous risk management are required. The impossibility of guaranteing security and robustness in light of the ever-growing attack surface due to IoT adoption slightly shifts the focus from security to resilience; from trying to ensure that a system is robust to trying to ensure that it can continue to operate, even when it has been partially compromised. This trend is further emphasised by the connection of the IoT to the physical world. System availability becomes more important as inter-dependencies across systems mean that disrupted operation can have significant cascading effects. The connection to the physical world also makes contextualisation a much more difficult challenge as systems must accommodate a variety of contexts of use, themselves evolving over time.     

The research landscape is heavily dominated by technologies and more importantly their adoption and acceptance. From trusted platforms such IntelSGX, to blockchain, AI, SDN and virtualisation, the application of new technologies forms the focus of many research investigations. Such technologies make it possible to implement old principles that have withstood the test of time such as mediation, isolation, detection, remediation, on which defences will continue to be based. Their adoption and use makes it possible sometimes to attempt to address some of the challenges specific to the IoT. 

However, this is not enough. Security and resilience are emergent properties and more so in a dynamic IoT systems than in enterprise systems and traditional networked environments. But these emergent properties need to arise from addressing inherently conflicting goals and requirements that are more stringent in IoT environments. Confidentiality and privacy can conflict with availability. Distribution and redundancy conflict with resource usage particularly in end devices, which often are very constrained in their energy consumption. Cost pressures and low margins make the deployment of hardware security solutions more difficult. Due to the pervasiveness of the IoT and its application across an infinite spectrum of applications and usages, it is difficult to characterise well the trade-offs underpinning these conflicts and develop more generic solutions, or at least patterns for such solutions. 

The usage of new technologies also has a side-effect: their vulnerability; they themselves are vulnerable to malicious attacks. This has been amply demonstrated with AI, with smart-contracts on the block-chain, with Intel SGX. The robustness of SDN to adversarial attacks remains under-researched. Ensuring the robustness of new technologies to adversarial attacks is sometimes far from trivial, the case of AI/Machine Learning standing as a particularly prominent example. The attack surface of Machine Learning algorithms and how to make them more robust remain fundamental research challenges. Yet, there is frequently more enthusiasm for the adoption of new technologies and their use towards ensuring security and resilience, than there is in ensuring their robustness. This is, at least in part, driven by the market pressures to innovate. 

However, progress is being made across all areas: in trusted platforms and secure hardware, in formal verification, in architectures that offer better mediation at a finer level of granularity (e.g., micro-segmentation), in leveraging redundancy for resilience. Many of the solutions developed for more traditional computing settings are being extended and transferred to the IoT domain. But in doing so the dynamcity and contextualisation of IoT systems prove to be a major challenge. This survey has collated current research into risk and threat mitigations in the IoT environment, providing an overview and presenting these uniquely  throughout the life cycle of the IoT device, starting from its design, to the moment when a device joins a network, while the IoT device operates in the network, and until it leaves (or is removed) from the network and is eventually decommissioned, hence adopting a “defence-in-depth” approach. The survey has discussed threat mitigation techniques applied across different IoT application contexts, and has elaborated on how to apply each mitigation technique, its benefits and limitations and the extent to which progress has been reported in the literature.

This survey has collated current research into risk and threat mitigations in the IoT environment, providing an overview and presenting these uniquely  throughout the life cycle of the IoT device, starting from its design, to the moment when a device joins a network, while the IoT device operates in the network, and until it leaves (or is removed) from the network and is eventually decommissioned, hence adopting a “defence-in-depth” approach. A summary of the discussed mitigation techniques is presented in Table \ref{tab: Table1}. The survey has discussed threat mitigation techniques applied across different IoT application contexts, and has elaborated on how to apply each mitigation technique, its benefits and limitations and the extent to which progress has been reported in the literature.

\begin{table}
    \centering
    \caption{A taxonomy for mitigation techniques throughout the life-cycle of an IoT device}
    \begin{tabular}{|m{4.5cm}|m{5cm}|m{3cm}|  }\hline
        IoT Device Life Cycle & Mitigation Technique &  Related Studies and Proposed work \newline \\\hline
        
        Before a device joins a system & Device self-protection and self-defence:\newline
        - Hardware self-protection \newline - Software self-protection \newline - Moving Target Defence  \newline &  \cite{jsan8030042, hamadeh2017area, lu2020xtseh, eldefrawy2012smart, mohan2018special,  9474036, 10.1145/2485922.2485970}  \newline \cite{1260985, 10.1007/978-3-030-64348-5_25, frank2018protecting, Choi2018SystemHA, ankergaard2021state}  \newline \cite{9521488, navas:hal-02887462, 9270287} \\\hline
        
       If a device wants to join a system  & Certification \newline &   \cite{10.1145/3410160, 10.1145/3410160}  \\\hline
        
        While the device is in the system & Mediation: \newline - IoT Edge \newline - Continuous monitoring: \newline
           *Intrusion detection \newline *Honeypots \newline *AI techniques for monitoring and detection \newline &   \cite{6996632, 7412116, 10.1145/2873587.2873600, 10.1145/3366612.3368122}  \newline  \cite{ZARPELAO201725}  \newline \cite{9520645, Vetterl2020HoneypotsIT, 191952} \newline \cite{kuzlu2021role, chaabouni2019network, kumar2021uids, meidan2018n, pacheco2019security, Pauna2019OnTR}  \\\hline
           
        When a cyber attack occurs& Device self-protection and self-defence:\newline
        - Hardware self-protection \newline - Software self-protection \newline - Moving Target Defence  \newline & \cite{jsan8030042, hamadeh2017area, lu2020xtseh, eldefrawy2012smart, mohan2018special,  9474036, 10.1145/2485922.2485970}  \newline \cite{1260985, 10.1007/978-3-030-64348-5_25, frank2018protecting, Choi2018SystemHA, ankergaard2021state} \newline \cite{9521488, navas:hal-02887462, 9270287} \\\hline
        
        If the device has been compromised & Device(s) isolation and system segmentation: \newline - 
        Identifying and documenting IoT dependencies \newline - Micro-Segmentation \newline - Software Defined Networks (SDN) \newline & \cite{stellios2018survey, 9174628, stergiopoulos2020automatic, luiijf2021analysis} \newline \cite{wasicek2020future, 253364, finproceedings} \newline \cite{baldini2020iot, garcia2019enforcing, zarca2019security} \\\hline
        
        When the device leaves or is removed from the system & System availability and resilience: \newline- Diversity \newline- Exploiting data correlation \newline - Exploiting data overhearing \newline - Network Coding (NC) \newline &  \cite{8539114, venkatakrishnan2016using} \newline \cite{salayma2017wireless, lupuillianosurvey, Illiano2017DontFM}  \newline \cite{9469421} \newline \cite{li2017towards, inbookt, 8664097, Lee2020AttackDU}  \\\hline
         
    \end{tabular}
    
    \label{tab: Table1}
\end{table}
\newpage
\begin{acks}
I would like to thank PETRAS National Centre of Excellence for IoT Systems Cybersecurity for funding this work \textit{https://petras-iot.org/}.
\end{acks}

\bibliographystyle{ACM-Reference-Format}
\bibliography{sample-base}

\end{document}